\documentclass[12pt]{article}

\usepackage{latexsym}
\usepackage{amssymb,amsfonts,amsmath}
\usepackage{graphicx} 
\usepackage{indentfirst}
\usepackage{bbm}
\usepackage{amssymb}
\usepackage{verbatim}
\usepackage{amsmath, amsthm,amssymb}
\usepackage{mathrsfs}
\usepackage{hyperref}
\usepackage{amsfonts}
\usepackage{dsfont}
\usepackage{cite}
\usepackage{xcolor}
\usepackage[multiple]{footmisc}

\parskip=\medskipamount

\newcommand {\cB}{{\cal B}}
\newcommand {\cC}{{\cal C}}
\newcommand {\cD}{{\cal D}}
\newcommand {\cE}{{\cal E}}

\newcommand {\cL}{{\cal L}}
\newcommand {\cM}{{\cal M}}
\newcommand {\cN}{{\cal N}}

\newcommand {\cW}{{\cal W}}

\def\a{\alpha}
\def \bi{\bibitem}

\def\b{\beta}
\def\c{\chi}
\def\d{\delta}

\def\g{\gamma}

\def\j{\psi}
\def\k{\kappa}
\def\l{\lambda}
\def\m{\mu}

\def\q{\theta}
\def\r{\rho}
\def\s{\sigma}

\def\z{\zeta}
\def\D{\Delta}
\def\F{\Phi}
\def\J{\Psi}

\def\U{\Upsilon}

\def\rd{{\rm d}}
\def\ri{{\rm i}}
\def\re{{\rm e}}

\newcommand{\ad}{{\dot{\alpha}}}                           
\newcommand{\bd}{{\dot{\beta}}}                            
\newcommand{\ve}{\varepsilon}                            
\newcommand{\cDB}{{\bar\cD}}                            

\renewcommand{\aa}{{\a\ad}}
\newcommand{\bb}{{\b\bd}}
\newcommand{\pa}{\partial}                           
\newcommand{\hf}{\frac12}

\newcommand{\be}{\begin{equation}}
\newcommand{\ee}{\end{equation}}
\newcommand{\bea}{\begin{eqnarray}}
\newcommand{\eea}{\end{eqnarray}}
\newcommand{\non}{\nonumber}

\newcommand{\2}{{\underline{2}}}

\newcommand{\bm}[1]{\mbox{\boldmath$#1$}}

\def\double #1{#1{\hbox{\kern-2pt $#1$}}}

\newcommand{\gd}{{\dot\g}}
\newcommand{\dd}{{\dot\d}}

\newcommand{\dalpha}{{\dot{\alpha}}}

\newcommand{\N}{{\mathcal N}}
\newif\ifdtup

\newcommand{\bsubeq}{\begin{subequations}}
\newcommand{\esubeq}{\end{subequations}}

\newcommand{\mub}{{{\bar{\mu}}}}

\newcommand{\eol}{\notag \\}

\numberwithin{equation}{section}

\newcommand{\sSU}{\mathsf{SU}}
\newcommand{\sSL}{\mathsf{SL}}

\newcommand{\sSO}{\mathsf{SO}}
\newcommand{\sU}{\mathsf{U}}

\begin{document}

\begin{titlepage}
\begin{flushright}
July, 2021 \\
Revised version: September, 2021\\
\end{flushright}
\vspace{5mm}

\begin{center}
{\Large \bf 
Duality-invariant (super)conformal higher-spin models}
\end{center}

\begin{center}

{\bf Sergei M. Kuzenko and Emmanouil S. N. Raptakis} \\
\vspace{5mm}

\footnotesize{
{\it Department of Physics M013, The University of Western Australia\\
35 Stirling Highway, Perth W.A. 6009, Australia}}  
~\\
\vspace{2mm}
~\\
Email: \texttt{ 
sergei.kuzenko@uwa.edu.au, emmanouil.raptakis@research.uwa.edu.au}\\
\vspace{2mm}

\end{center}

\begin{abstract}
\baselineskip=14pt
We develop a general formalism of duality rotations for bosonic conformal spin-$s$ gauge fields, with $s\geq 2$, in a conformally flat four-dimensional spacetime. In the $s=1$ case this formalism is equivalent to the theory of $\mathsf{U}(1)$ duality-invariant nonlinear electrodynamics developed by  Gaillard and Zumino, Gibbons and Rasheed, and generalised by Ivanov and Zupnik. For each integer spin $s\geq 2$ we demonstrate the existence of families of conformal  $\mathsf{U}(1)$ duality-invariant models, including a generalisation of the so called ModMax Electrodynamics ($s=1$).
Our bosonic results are then extended to the $\cN=1$ and $\cN=2$ supersymmetric cases. 
We also sketch a formalism of duality rotations for conformal gauge fields of Lorentz type
$(m/2, n/2)$, for positive integers $m $ and $n$.
\end{abstract}
\vspace{5mm}

\vfill

\vfill
\end{titlepage}

\newpage
\renewcommand{\thefootnote}{\arabic{footnote}}
\setcounter{footnote}{0}

\tableofcontents{}
\vspace{1cm}
\bigskip\hrule

\allowdisplaybreaks


\section{Introduction}

Building on the seminal 1981 work by Gaillard and Zumino \cite{GZ1},
the general theory of $\sU(1)$ duality-invariant
models for nonlinear electrodynamics in four dimensions was developed in the mid 1990s \cite{GR1,GR2,GZ2,GZ3} and the early 2000s \cite{IZ_N3,IZ1,IZ2}.
The formalism of  \cite{GR1,GR2,GZ2,GZ3} 
 has been generalised to formulate general 
$\sU(1)$ duality-invariant $\cN=1$ and $\cN=2$ globally \cite{KT1,KT2} and locally \cite{KMcC,KMcC2,K12} supersymmetric theories. 
In particular, Ref. \cite{KT1} put forward the constructive perturbative scheme to compute $\cN=2$ superconformal 
$\sU(1)$ duality-invariant actions for the $\cN=2$ vector multiplet.
Moreover, extending the earlier proposal of \cite{Ketov}, the first consistent 
perturbative scheme to construct the $\cN=2$ supersymmetric Born-Infeld action 
was given in \cite{KT2}. The formalism of nonlinear realisations for
the partial  $\cN=4 \to \cN=2$ breaking of supersymmetry advocated in \cite{BIK1}
reproduced \cite{BIK2} the results of \cite{KT2}.
Further progress toward the construction of the $\cN=2$ supersymmetric Born-Infeld action has been achieved in \cite{BCFKR,CK,IZ4}.\footnote{It should be pointed out that the $\cN=1$ supersymmetric Born-Infeld action \cite{CF} is the first nontrivial $\sU(1)$ duality-invariant supersymmetric theory \cite{BMZ}. Its remarkable property is that it is a Goldstone multiplet 
action for partial   $\cN=2 \to \cN=1$ supersymmetry 
breaking in Minkowski space \cite{BG,RT},  as well as in the maximally supersymmetric backgrounds \cite{KT-M16} discovered in \cite{FS}: 
(i)  ${\mathbb R} \times S^3$; (ii) ${\rm AdS}_3 \times {\mathbb R}$;
and (iii) a supersymmetric plane wave.  }

Within the original bosonic formulation \cite{GR1,GR2,GZ2,GZ3} and its supersymmetric extensions \cite{KT1,KT2}, $\sU(1)$ duality invariance of a model for nonlinear (super) electrodynamics is equivalent to the condition that the Lagrangian satisfies a nonlinear self-duality equation. General solutions of such equations are difficult to find.
Ivanov and Zupnik \cite{IZ_N3,IZ1,IZ2} provided a reformulation of nonlinear electrodynamics which makes use of certain auxiliary variables in such a way that 
 (i) the self-interaction depends only on the auxiliary variables; and (ii) 
  $\sU(1)$ duality invariance is equivalent to the manifest $\sU(1)$ invariance of the self-interaction. Supersymmetric extensions of the Ivanov-Zupnik approach were given in 
\cite{K13,ILZ}. In particular, the $\cN=2$ supersymmetric formulation of \cite{K13}
has been used to obtain the closed-form expression for a superconformal $\sU(1)$ duality-invariant model proposed to describe the low-energy effective action for $\cN=4$ super-Yang-Mills theory \cite{K21}.

In this paper we will demonstrate that the known results for $\sU(1)$ duality-invariant nonlinear electrodynamics (spin $s=1$) 
can naturally be generalised  to develop a general formalism of $\mathsf{U}(1)$ duality rotations for bosonic conformal spin-$s$ gauge fields, with $s\geq 2$, 
and their $\cN=1$ and $\cN=2$ supersymmetric cousins 
in a conformally flat four-dimensional background. 

Our paper is organised as follows. In section \ref{section2} we introduce the notion of $\sU(1)$ duality-invariant conformal higher-spin (CHS) theories and present examples of such models, including higher-spin generalisations of the so-called ``ModMax electrodynamics" \cite{BLST} (see also \cite{Kosyakov}). 
This purely bosonic study is extended in section \ref{section3} to the case of  $\cN=1$ superconformal higher-spin (SCHS) multiplets. We present a one-parameter self-dual SCHS action, which generalises the 
$\cN=1$ superconformal $\sU(1)$ duality-invariant electrodynamics 
\cite{BLST2,K21}. In section \ref{section4} we uplift the technical machinery of the previous two sections to $\N=2$ superspace and derive the self-duality equation for $\N=2$ SCHS models. 
In section \ref{section5} we provide concluding comments and sketch the formalism
of duality rotations for CHS gauge fields of arbitrary rank.  
The main body of this paper is accompanied by four technical appendices. 
Appendix  \ref{appendixA} reviews the salient details of conformal geometry in four dimensions pertinent to this work.
Appendices \ref{appendixB} and \ref{appendixC} review the relevant aspects of $\N =1$ and 
$\cN=2$   conformal superspace, respectively. 
In appendix \ref{appendixNew} we derive a class of duality-invariant CHS models 
via the auxiliary field approach. 
In appendix \ref{appendixD} we provide arguments to fix the overall signs of the free (super)conformal higher-spin actions.


\section{Duality-invariant CHS models} \label{section2}

In this section we develop a formalism of duality rotations for CHS fields and propose some duality-invariant models.

Consider a dynamical system describing the propagation of a conformal spin-$s$ field
$h_{\a(s) \ad(s)}: = h_{\a_1 \dots \a_s \ad_1 \dots \ad_s} 
=h_{(\a_1 \dots \a_s)( \ad_1 \dots \ad_s) }$,  with $s\geq 1$, 
 in curved spacetime. Its action functional 
$S^{(s)}[\mathcal{C},\bar{\mathcal C}]$ is assumed to depend 
on a field strength $\mathcal{C}_{\a(2s)}$ and its conjugate 
$\bar{\mathcal{C} }_{\ad (2s) }$, with $\mathcal{C}_{\a(2s)}$ being defined as 
\be
\label{2.1}
\mathcal{C}_{\a(2s)} =  \nabla_{(\a_1}{}^{\bd_1} \dots \nabla_{\a_s}{}^{\bd_s} h_{\a_{s+1} \dots \a_{2s}) \bd(s)} ~, 
\ee
where $\nabla_a$ denotes the conformally covariant derivative, see appendix \ref{appendixA}. 
The real unconstrained prepotential $h_{\a(s) \ad(s)}$ is a primary field, 
$K_b h_{\a(s) \ad(s)} =0$, where $K_b$ is the special conformal generator. 
It is
defined modulo gauge transformations of the form
\be
\label{2.2}
\d_\z h_{\a(s) \ad(s)} = \nabla_{(\a_1 (\ad_1} \z_{\a_2 \dots \a_s) \ad_2 \dots \ad_s)}~,
\ee
with the gauge parameter $\z_{\a(s-1) \ad(s-1)}$ also being  primary.
This transformation law is conformally invariant provided
\be
\mathbb{D} h_{\a(s) \ad(s)} = (2 - s) h_{\a(s) \ad(s)} 
~,
\ee
where $\mathbb D$ is the dilatation generator.
The field strength \eqref{2.1} is primary in a generic gravitational background, 
\bea
K_b \mathcal{C}_{\a(2s)} =0~, \qquad {\mathbb D} \mathcal{C}_{\a(2s)} = 2 \mathcal{C}_{\a(2s)}~.
\eea
However, 
the gauge transformations \eqref{2.2} leave $\mathcal{C}_{\a(2s)}$ invariant only when $s=1$, $\d_\z \mathcal{C}_{\a(2)} = 0$. For $s \geq 2$ gauge invariance holds only if the background is conformally flat, 
\bea
C_{\a(4)}=0 \quad \implies \quad \d_\z \mathcal{C}_{\a(2s)} = 0~,
\eea
where $C_{\a(4)}$ is the self-dual part of the background Weyl tensor, 
see appendix \ref{appendixA}.
For the remainder of this section we will assume such a geometry.

We point out that $\mathcal{C}_{\a(2)}$ is Maxwell's field strength and $\mathcal{C}_{\a(4)}$ is the linearised Weyl tensor. We will refer to  $\mathcal{C}_{\a(2s)}$ for $s>2$ as the linearised spin-$s$ Weyl tensor.


\subsection{$\sU(1)$ duality-invariant models}

It is important to note that the field strength \eqref{2.1} obeys the Bianchi identity
\be
\label{2.4}
\nabla^{\b_1}{}_{(\ad_1} \dots \nabla^{\b_s}{}_{\ad_s)} \mathcal{C}_{\a(s) \b(s)} 
=
\nabla_{(\a_1}{}^{\bd_1} \dots \nabla_{\a_s)}{}^{\bd_s} \bar{\mathcal{C}}_{\ad(s) \bd(s)}  ~.
\ee
Now, we assume that  $S^{(s)}[\mathcal{C},\bar{\mathcal C}]$ is extended  to be a functional of an unconstrained field $\mathcal{C}_{\a(2s)}$ and its conjugate. We introduce
\be
\ri \mathcal{M}_{\a(2s)} := \frac{\d S^{(s)}[\mathcal{C},\bar{\mathcal{C}}]}{\d \mathcal{C}^{\a(2s)}} ~,
\ee
where we have defined
\be
\d S^{(s)}[\mathcal{C},\bar{\mathcal{C}}] = \int \rd^4x\, e \, \d \mathcal{C}^{\a(2s)} \frac{\d S^{(s)}[\mathcal{C},\bar{\mathcal{C}}]}{\d \mathcal{C}^{\a(2s)}}  + \text{c.c.}
\ee
Varying $S^{(s)}[\mathcal C , \bar{\mathcal C} ]$ with respect to the prepotential $h_{\a(s) \ad(s)}$ yields
\be
\label{2.8}
\nabla^{\b_1}{}_{(\ad_1} \dots \nabla^{\b_s}{}_{\ad_s)} \mathcal{M}_{\a(s) \b(s)} 
=
\nabla_{(\a_1}{}^{\bd_1} \dots \nabla_{\a_s)}{}^{\bd_s} \bar{\mathcal{M}}_{\ad(s) \bd(s)}  ~.
\ee

A crucial feature of our analysis above is that the functional form of the equation of motion \eqref{2.8} mirrors that of the Bianchi identity \eqref{2.4}. Consequently,
the union of equations  \eqref{2.4} and  \eqref{2.8}
is invariant under the $\sSO(2)\cong \sU(1)$ duality transformations:
\be
\label{2.9}
\d_\l \mathcal{C}_{\a(2s)} = \l \mathcal{M}_{\a(2s)} ~, \quad \d_\l \mathcal{M}_{\a(2s)} = - \l \mathcal{C}_{\a(2s)} ~,
\ee
where $\l$ is a constant, real parameter. One may then obtain two equivalent expressions for the variation of $S^{(s)}[\mathcal C , \bar{\mathcal C}]$ with respect to \eqref{2.9}
\be
\label{U1variation}
\d_\l S^{(s)}[\cC,\bar{\cC}] = \frac{\ri \l}{4} \int \rd^4x \, e \, \Big \{ \cC^2 - \cM^2 \Big \} + \text{c.c.} = -\frac{\ri \l}{2} \int \rd^4x \, e \, \cM^2 + \text{c.c.}~,
\ee
as a generalisation of similar derivations in nonlinear electrodynamics 
\cite{GZ2,GZ3,KT2}. 
This implies the self-duality equation
\be
\label{2.10}
\text{Im} \int \rd^4x \, e \, \Big \{ \mathcal{C}^{\a(2s)}  \mathcal{C}_{\a(2s)}
+ \mathcal{M}^{\a(2s)} \mathcal{M}_{\a(2s)} \Big \}  = 0 ~,
\ee
which must hold for an unconstrained field $\mathcal{C}_{\a(2s)}$ and its conjugate. 
In \eqref{U1variation} we have employed the notational shorthand $T^2 = T^{\a(m)} T_{\a(m)}$ (similarly $\bar{T}^2 = \bar{T}_{\ad(m)} \bar{T}^{\ad(m)}$).
The simplest solution of the self-duality equation \eqref{2.10} is the
free CHS model \eqref{D.1}, which was introduced 
in \cite{FT,FL,FL2} in the case of Minkowski space and extended to arbitrary conformally flat backgrounds in \cite{KP}.\footnote{For the free CHS 
model \eqref{D.1}, one can also consider scale transformations in addition to 
the $\sU(1)$ duality ones \eqref{2.9}, which is similar to the case of electrodynamics
discussed, e.g., in \cite{KT2}.}

In the $s=1$ case, the self-duality equation \eqref{2.10} was originally 
derived by Bialynicki-Birula \cite{B-B}, but unfortunately this work was largely unnoticed.\footnote{We thank Dmitri Sorokin for bringing Ref. \cite{B-B} to our attention.}
It was independently re-discovered  
by Gibbons and Rasheed in 1995 \cite{GR1}.
Two years later, it was re-derived by  Gaillard and Zumino  \cite{GZ2} with the aid of their formalism 
developed back in 1981 \cite{GZ1} but originally applied only in the linear case.

As is known, all gravity-matter theories allow for a Weyl-invariant formulation  \cite{Deser, Zumino} in which the gravitational field is described in terms of two gauge fields. 
One of them is the inverse vielbein $e_a{}^m $ and the other is 
a conformal compensator $\J$, the latter being a nowhere vanishing scalar field. 
In this setting the gravity gauge group  also includes
Weyl transformations, which act
on the gravitational fields as follows
\bea 
\label{Weyl}
 e_a{}^m \to \re^\s e_a{}^m ~, 
\qquad \J \to  \re^{\s} \J~. 
\eea
Truly conformal theories, such as conformal gravity, do not depend on the compensator. In the approach of \cite{KakuT}, the gauge group is further enlarged to local conformal transformations, and the compensator is a primary dimension-1 scalar field, 
\bea
K_b \J =0~, \qquad {\mathbb D} \J = \J~.
\eea

If we allow for the action  $S^{(s)}[\mathcal{C},\bar{\mathcal C}]$  to depend on the compensator, 
then the  family of $\sU(1)$ duality-invariant theories \eqref{2.10}
 is very large. 
For instance, the following $\sU(1)$ duality-invariant model 
\bea
\label{HSBI}
S^{(s)}_\text{BI}[\cC,\bar{\cC} ; \Psi] = - \int \rd^4x \, e \, \J^4 \bigg \{ 1 - \bigg (1 + (-1)^s \frac{\cC^2 + \bar{\cC}^2}{\J^4} + \frac{(\cC^2 - \bar \cC^2)^2}{4 \J^8} \bigg )^{\frac{1}{2}} \bigg \} 
\eea
is a higher-spin generalisation of Born-Infeld electrodynamics \cite{BI}. 
The latter is obtained from \eqref{HSBI} for $s=1$ by making use of local scale transformations to impose a gauge condition $\J^2 = g^{-1} = {\rm const}$.
Owing to the dependence of $S^{(s)}_\text{BI}[\cC,\bar{\cC} ; \Psi] $ on the compensator $\J$, it is clear that \eqref{HSBI} is not conformal.

As another solution of the self-duality equation \eqref{2.10}, we propose a one-parameter duality-invariant extension of \eqref{HSBI}
\begin{align}
\label{HSBIgen}
S^{(s)}_\text{BIgen}[\cC,\bar{\cC} ; \Psi] &= - \int \rd^4x \, e \, \J^4 \bigg \{ 1 - \bigg (  1 + \frac{2}{\Psi^4} \bigg [ \frac{(-1)^s}{2} \text{cosh} \g (\cC^2 + \bar{\cC}^2) + \sinh \g (\cC^2 \bar{\cC}^2)^{\frac 1 2} \bigg ] \non \\
& \qquad \qquad \qquad \qquad + \frac{(\cC^2 - \bar \cC^2)^2}{4 \J^8} \bigg)^{\frac 1 2} \bigg \}~, \qquad \g \in \mathbb{R}~.
\end{align}
For $s=1$ this model was introduced in \cite{Bandos:2020hgy}.


\subsection{Self-duality under Legendre transformation} \label{section2.2}

In the case of nonlinear (super) electrodynamics,  $\sU(1)$ duality invariance implies self-duality under Legendre transformations, see \cite{KT2} for a review. This remarkable property proves to extend to the higher-spin case, as will be shown below.

We start by describing a Legendre transformation for a generic theory with action 
 $S^{(s)}[\mathcal{C},\bar{\mathcal C}]$.
For this we introduce the parent action
\begin{align}
	\label{parent}
	S^{(s)} [\mathcal C,\bar{\mathcal C},
	\mathcal C^{\rm D},\bar{\mathcal C}^{\rm D}] = S^{(s)}[\mathcal C,\bar{\mathcal C}] + \int \rd^4x \, e \, \Big ( \frac \ri 2 \mathcal C^{\a(2s)} \mathcal C^{\rm D}_{\a(2s)} + \text{c.c.} \Big ) ~.
\end{align}
Here $\mathcal{C}_{\a(2s)}$ is an unconstrained field and $\mathcal C^{\rm D}_{\a(2s)}$ takes the form
\begin{align}
	\mathcal C^{\rm D}_{\a(2s)} = \nabla_{(\a_1}{}^{\bd_1} \dots \nabla_{\a_s}{}^{\bd_s} h^{\rm D}_{\a_{s+1} \dots \a_{2s}) \bd(s)} ~,
\end{align}
where $h^{\rm D}_{\a(s) \ad(s)}$ is a Lagrange multiplier field. Indeed, upon varying \eqref{parent} with respect to 
 $h^{\rm D}_{\a(s) \ad(s)}$ 
one obtains the Bianchi identity \eqref{2.4}, 
and its general solution is given by eq. \eqref{2.1}, for some real field $h_{\a(s)\ad(s)}$. 
As a result the second term in \eqref{parent} becomes a total derivative, and we end up with the original action 
 $S^{(s)}[\mathcal{C},\bar{\mathcal C}]$.
Alternatively, if we first vary  \eqref{parent} with respect to $\mathcal{C}^{\a(2s)}$, the equation of motion is
\begin{align}
	\mathcal{M}_{\a(2s)} = - \mathcal{C}^{\rm D}_{\a(2s)}~,
\end{align}
which we may solve to express $\mathcal{C}_{\a(2s)}$ as a function of 
$\mathcal{C}^{\rm D}_{\a(2s)}$ and its conjugate. 
Inserting this solution into \eqref{parent}, we obtain the dual model
\begin{align}
	\label{dualmodel}
	S^{(s)}_{\rm D}[\mathcal C^{\rm D},\bar{\mathcal C}^{\rm D}] 
	:= \Big [ S^{(s)}[\mathcal C,\bar{\mathcal C}] +  \int \rd^4x \, e \, \Big ( \frac \ri 2 \mathcal C^{\a(2s)} \mathcal C^{\rm D}_{\a(2s)} + \text{c.c.} \Big ) \Big ]\Big |_{\mathcal{C} = \mathcal{C}( \mathcal{C}^{\rm D},  \bar{\mathcal{C}}^{\rm D})}~.
\end{align}

Now, given an action $S^{(s)}[\mathcal C , \bar{\mathcal C} ]$ satisfying \eqref{2.10}, our aim is to show that it satisfies
\begin{align}
	\label{legendre}
	S^{(s)}_{\rm D}[\mathcal C,\bar{\mathcal C}] = S^{(s)} [\mathcal C,\bar{\mathcal C}]~,
\end{align}
which means that the corresponding Lagrangian is invariant under Legendre transformations. A routine calculation allows one to show that the following functional
\begin{align}
	\label{invariant}
	S^{(s)}[\mathcal C,\bar{\mathcal C}] +  \int \rd^4x \, e \, \Big ( \frac \ri 4 \mathcal C^{\a(2s)} \mathcal M_{\a(2s)} + \text{c.c.} \Big )
\end{align}
is invariant under \eqref{2.9}. The latter may be exponentiated to obtain the finite $\sU(1)$ duality transformations
\begin{align}
	\mathcal{C}'_{\a(2s)} = \text{cos} \l \, \mathcal{C}_{\a(2s)} + \text{sin} \l \, \mathcal{M}_{\a(2s)} ~, \quad
	\mathcal{M}'_{\a(2s)} = - \text{sin} \l \, \mathcal{C}_{\a(2s)} + \text{cos} \l \, \mathcal{M}_{\a(2s)}  ~.
\end{align}
Performing such a transformation with $\l = \frac \pi 2$ on \eqref{invariant} yields 
\begin{align}
	S^{(s)}[\mathcal C, \bar{\mathcal C}] 
	= S^{(s)}[\mathcal C^{\rm D}, \bar{\mathcal C}^{\rm D}] 
	- \int \rd^4x \, e \, \Big( \frac \ri 2 \mathcal{C}^{\a(2s)} \mathcal{C}^{\rm D}_{\a(2s)} + \text{c.c.} \Big) ~. 
\end{align}
Upon inserting this expression into \eqref{dualmodel}, we obtain \eqref{legendre}.

In the above analysis, we made use of the fact that the  general solution of  the Bianchi identity \eqref{2.4} is given by \eqref{2.1}. To justify this claim, it suffices to work in Minkowski space. Let $\cC_{\a(2s)} $ be a field subject to the equation \eqref{2.4}, with $\nabla_a=\pa_a$. Introduce its descendant defined by
\begin{subequations}
\bea
h^{\perp}_{\a(s) \ad(s)} := \pa^{\b_1}{}_{\ad_1} \dots \pa^{\b_s}{}_{\ad_s} \mathcal{C}_{\a(s) \b(s)} &=&
\pa^{\b_1}{}_{(\ad_1} \dots \pa^{\b_s}{}_{\ad_s)} \mathcal{C}_{\a(s) \b(s)} ~,
\label{perp}
\eea
which is automatically transverse, 
\bea
\pa^{\b\bd} h^{\perp}_{\b\a(s-1) \bd \ad(s-1)} &=&0~.
\eea
\end{subequations}
The Bianchi identity \eqref{2.4} tells us that $h^{\perp}_{\a(s) \ad(s)} $ is real, 
$\overline{h^{\perp}_{\a(s) \ad(s)}  }= h^{\perp}_{\a(s) \ad(s)} $.
Now we can express $\cC_{\a(2s)} $ in terms of \eqref{perp},
\bea
\mathcal{C}_{\a(2s)} =  \Box^{-s} \pa_{(\a_1}{}^{\bd_1} \dots \pa_{\a_s}{}^{\bd_s} 
h^{\perp}_{\a_{s+1} \dots \a_{2s}) \bd(s)} 
=\pa_{(\a_1}{}^{\bd_1} \dots \pa_{\a_s}{}^{\bd_s} 
h_{\a_{s+1} \dots \a_{2s}) \bd(s)} ~,
\eea
where $\Box = \pa^a \pa_a$.
In the final relation the real field $h_{\a(s)\ad(s)}$ is not assumed to be transverse.
This field is related to $\Box^{-s} h^{\perp}_{\a (s) \ad(s)} $ by a gauge transformation
\bea
\d_\z h_{\a(s) \ad(s)} = \pa_{(\a_1 (\ad_1} \z_{\a_2 \dots \a_s) \ad_2 \dots \ad_s)}~,
\eea
with a real gauge parameter $\z_{\a(s-1) \ad(s-1)}$.  Our consideration may be extended to the $\cN=1$ and $\cN=2$ supersymmetric cases studied in the next sections.


\subsection{Auxiliary variable formulation} \label{section2.3}

As a generalisation of the Ivanov-Zupnik \cite{IZ_N3,IZ1,IZ2} approach, 
here we will introduce a powerful formalism to generate duality-invariant models that makes use of auxiliary variables. 

Consider the following action functional
\begin{align}
	\label{auxiliaryCHS}
	S^{(s)}[\mathcal{C},\bar{\mathcal C}, \r, \bar \r] &= (-1)^s \int \rd^4x 
	\, e \, \Big \{ 2 \r \mathcal{C} - 
	\r^2- \frac{1}{2}\mathcal{C}^2 \Big \} + \text{c.c.} + \mathcal{S}^{(s)}_{\text{int}} [\r , \bar{\r}] ~.
\end{align}
Here we have introduced the auxiliary variable $\r_{\a(2s)}$
which is chosen to be an unconstrained primary dimension-2 field, 
\bea
K_b \mathcal{\r}_{\a(2s)} =0~, \qquad {\mathbb D} \mathcal{\r}_{\a(2s)} = 2 \mathcal{\r}_{\a(2s)}~.
\eea
The functional 
$\mathcal{S}^{(s)}_{\text{int}} [\r , \bar{\r}]$, by definition, contains cubic and higher powers of $\r_{\a(2s)}$ and its conjugate. The equation of motion for $\r^{\a(2s)}$ is 
\be
\label{etaEoM}
\r_{\a(2s)} = \mathcal{C}_{\a(2s)} + \frac{(-1)^s}{2} \frac{\d \mathcal{S}^{(s)}_{\text{int}} [\r , \bar{\r}]}{\d \r^{\a(2s)}}~.
\ee
Equation \eqref{etaEoM} allows one to express $\r_{\a(2s)}$ as a functional of $\mathcal{C}_{\a(2s)}$ and its conjugate. This means that \eqref{auxiliaryCHS} is equivalent to a CHS theory with action
\begin{align}
	\label{dualTheory}
	S^{(s)}[\mathcal{C},\bar{\mathcal C}] &= \frac{(-1)^s}{2} \int \rd^4x 
	\, e \, \mathcal{C}^2 + \text{c.c.} + {S}^{(s)}_{\text{int}} [\mathcal{C} , \bar{\mathcal C}] ~.
\end{align}
Thus, \eqref{auxiliaryCHS} and \eqref{dualTheory} provide two equivalent realisations of the same model.

The power of this formulation is most evident when the self-duality equation \eqref{2.10} is applied. A routine computation reveals that this constraint is equivalent to
\be
\label{dualCondition}
\text{Im} \int \rd^4x \, e \, \r^{\a(2s)} \frac{\d \mathcal{S}^{(s)}_{\text{int}} [\r , \bar{\r}]}{\d \r^{\a(2s)}} = 0 ~.
\ee
Thus, self-duality of the action \eqref{auxiliaryCHS} is equivalent to the requirement that $\mathcal{S}^{(s)}_{\text{int}}[\r,\bar{\r}]$ is invariant under rigid $\sU(1)$ phase transformations
\be
\mathcal{S}^{(s)}_{\text{int}} [\re^{\ri \varphi} \r , \re^{- i \varphi} \bar{\r}] 
= \mathcal{S}^{(s)}_{\text{int}} [\r , \bar{\r}] ~, \quad \varphi \in \mathbb{R} ~.
\label{dualCondition2}
\ee

 For instance we can consider the model 
 \bea 
  \mathcal{S}^{(s)}_{\text{int}} [\r , \bar{\r} ; \Psi]  = 
  \int \rd^4x \, e \, \J^4 {\mathfrak F} \Big( 
  \frac{ \mathcal{\r}^2 \bar{\mathcal{\r}}^2 }{\J^8} \Big)~,
\eea
where ${\mathfrak F}(x) $ is a real analytic function of a real variable. However, such models are not conformal if the action does depend on $\J$. The condition of conformal invariance imposes additional nontrivial restrictions. 


\subsection{Conformal $\sU(1)$ duality-invariant models} \label{section2.4}

In the $s=1$ case, there is  the unique conformal $\sU(1)$ duality-invariant electrodynamics proposed in \cite{BLST} (see also \cite{Kosyakov}). It was called  ``ModMax electrodynamics'' in \cite{BLST}.
It turns out that for $s>1$, families of conformal $\sU(1)$ duality-invariant models exist.

As a warm-up example, let us consider the following nonlinear conformal action
\begin{align}
	\label{2.11}
	S^{(s)}[\mathcal{C},\bar{\mathcal C}] = \frac{ (-1)^s \a}{2} \int \rd^4x \, e \, \Big \{ \mathcal{C}^2 + \bar{\mathcal{C}}^2 \Big \} + \b \int \rd^4x \, e \, \sqrt{\mathcal{C}^2 \bar{\mathcal{C}}^2} ~, \quad \a,\b \in \mathbb{R}.
\end{align}
Requiring this action to obey the self-duality equation \eqref{2.10}, we obtain the constraint
\be
\a^2 - \b^2 = 1 \quad \Longrightarrow \quad \a = \text{cosh} \, \g ~, \quad \b = \text{sinh} \, \g ~, \quad \g \in \mathbb{R}~.
\ee
Thus, the nonlinear theory
\begin{align}
	\label{2.13}
	S^{(s)}[\mathcal{C},\bar{\mathcal C}] = \frac{(-1)^s \text{cosh} \, \g}{2} \int \rd^4x \, e \, \Big \{ \mathcal{C}^2 + \bar{\mathcal{C}}^2 \Big \} + \text{sinh} \, \g \int \rd^4x \, e \, \sqrt{\mathcal{C}^2 \bar{\mathcal{C}}^2} ~, \quad \g \in \mathbb{R},
\end{align}
is a one-parameter conformal $\sU(1)$ duality-invariant  extension of the free CHS action \eqref{D.1}.
In the $s=1$ case our model coincides with  ModMax electrodynamics.

In order to construct more general models, it is advantageous to make use of the auxiliary variable formulation described above. We introduce algebraic invariants of the symmetric rank-$(2s)$ spinor $\r_{\a(2s)} $, which has the same algebraic properties as  the linearised spin-$s$ Weyl tensor $\cC_{\a(2s)}$:
\bea
\r^2 := (-1)^s \r_{\a(s)}{}^{\b(s)} \r_{\b(s)}{}^{\a(s)} ~, \qquad 
\r^3 := \r_{\a(s)}{}^{\b(s)} \r_{\b(s)}{}^{\g(s)} \r_{\g(s)}{}^{\a(s)}~, \qquad \dots
\label{invariants}
\eea
If $s$ is odd, all invariants $\r^{2n +1}$, with $n$ a non-negative integer, vanish.

For simplicity, we restrict our analysis to the conformal graviton, $s=2$. 
In this case the family of invariants \eqref{invariants} contains only two functionally independent invariants \cite{PenroseR}, $\r^2 $ and $\r^3$. In particular, one may show that 
\bea
s=2: \qquad \r^4 = \hf (\r^2)^2~.
\eea
Now we choose the self-interaction in \eqref{auxiliaryCHS} to be of the form 
\bea
\label{2.35}
 \mathcal{S}^{(2)}_{\text{int}} [\r , \bar{\r}] 
=\int \rd^4 x\, e\, \Big\{ \b \big( \r^2 \bar \r^2\big)^{\hf} + \k \big( \r^3 \bar \r^3\big)^{\frac 13} \Big\}~,
\eea
where $\b$ and $\k$ are real coupling constants. The resulting model is clearly 
conformal and $\sU(1)$ duality-invariant. For $\k \neq 0$, elimination of the auxiliary variables $\r_{\a(4)}$ and $\bar \r_{\ad(4)}$ does not 
result 
in a simple action like \eqref{2.13}. In particular, such an elimination, to quadratic order in the couplings, yields the following self-dual model
\bea
\label{confgravitonNLAction}
S^{(2)}[\cC,\bar{\cC}] &=& \int \rd^4 x\, e\, \bigg\{ 
\frac 1 2 \Big (1 + \frac 1 2 \b^2 \Big) (\cC^2 +\bar \cC^2) 
+ \b (\cC^2 \bar{\cC}^2)^{\frac 1 2} + \kappa (\cC^3 \bar{\cC}^3)^{\frac 1 3} \non \\
&& \quad + \frac{1}{2} \b \k \frac{(\cC^3)^2 \bar{\cC}^2 + (\bar{\cC^3})^2 {\cC}^2 }{(\cC^3 \bar{\cC}^3)^{\frac 2 3} (\cC^2 \bar{\cC}^2)^{\frac 1 2}}
+ \frac{1}{12} \kappa^2 \frac{(\cC^2)^2 + (\bar{\cC}^2)^2}{(\cC^3 \bar{\cC}^3)^{\frac 1 3}} \non \\
&& \quad - \frac{1}{24} \kappa^2 \frac{(\cC^3)^2 (\bar{\cC}^2)^2 + (\bar{\cC^3})^2 ({\cC}^2)^2}{(\cC^3 \bar{\cC}^3)^{\frac 4 3}} + \dots \bigg \} ~.
\eea
The ellipsis in \eqref{confgravitonNLAction} denotes additional contributions to the full nonlinear theory which are cubic or higher order in the coupling constants.
We emphasise that for the special case $\k=0$ the above action yields 
\eqref{2.13}. A proof of this result is given in appendix \ref{appendixNew}.

For $s> 2$ the number of algebraic invariants of $\r_{\a(2s)}$ grows, and therefore one can define families of conformal $\sU(1)$ duality-invariant models.


\section{$\mathcal{N} = 1$ duality-invariant SCHS models} \label{section3}

The purely bosonic study undertaken in the previous section can be generalised to the supersymmetric case. To this end, we consider a dynamical system describing the propagation of a conformal superspin-$(s+\hf)$ gauge multiplet $H_{\a(s) \ad(s)}$, $s > 0$, 
in $\cN=1$ curved superspace  \cite{KMT,KP}.
The prepotential $H_{\a(s) \ad(s)}$ is a real unconstrained superfield being defined modulo the gauge transformations
\be
\label{3.2}
\d_\z H_{\a(s) \ad(s)} = \nabla_{(\a_1} \bar{\z}_{\a_2 \dots \a_s) \ad(s)} - \bar{\nabla}_{(\ad_1} \z_{\a(s) \ad_2 \dots \ad_s)}~,
\ee
with a primary unconstrained gauge parameter $\z_{\a(s) \ad(s-1)}$.
This gauge transformation law is superconformal provided
\be
K^B H_{\a(s)\ad(s)} =0~, \qquad 
\mathbb{D} H_{\a(s) \ad(s)} = - s H_{\a(s) \ad(s)} 
\ee
where  $K^B=(K^b, S^\b ,\bar S_\bd)$ are the special conformal generators, 
see appendix \ref{appendixB}.
The action functional of the dynamical system is required to take the form $S^{(s)}[\mathcal{W},\bar{\mathcal W}]$, where $\mathcal{W}_{\a(2s+1)}$
is the following chiral descendant of $H_{\a(s) \ad(s)}$:
\be
\label{3.1}
\mathcal{W}_{\a(2s+1)} = - \frac{1}{4} \bar{\nabla}^2 \nabla_{(\a_1}{}^{\bd_1} \dots \nabla_{\a_s}{}^{\bd_s} \nabla_{\a_{s+1}} H_{\a_{s+2} \dots \a_{2s+1}) \bd(s)} ~, 
\qquad \bar \nabla^\bd \mathcal{W}_{\a(2s+1)} = 0~.
\ee
This field strength proves to be primary in a generic supergravity background \cite{KMT,KP},
\bea
K^B \cW_{\a(2s+1)} =0~, \qquad 
\mathbb{D} \mathcal{W}_{\a(2s+1)} = \frac 3 2 \mathcal{W}_{\a(2s+1)}~.
\eea
However, the gauge transformations \eqref{3.2} leave $\mathcal{W}_{\a(2s+1)}$ invariant 
only if the supergravity background is conformally flat, 
\bea
W_{\a(3)}=0 \quad \implies \quad \d_\z \mathcal{W}_{\a(2s+1)} = 0~,
\eea
where $W_{\a(3)}$ is the background super-Weyl tensor, see appendix \ref{appendixB}.
For the remainder of this section we will assume such a geometry.

It should be pointed out that the prepotential $H_\aa$ encodes the linearised conformal supergravity multiplet and its field strength $\cW_{\a(3)}$ is the linearised 
super-Weyl tensor \cite{OS,FZ2}. We will refer to $\mathcal{W}_{\a(2s+1)} $ for $s>1$ as the linearised 
superspin-$(s+\hf)$ Weyl tensor, or simply as a
higher-spin super-Weyl tensor. 

With the exception of the gauge transformation law \eqref{3.2}, the above results are also valid in the $s=0$ case corresponding to the massless vector multiplet.  As is well known, its prepotential $H$ is defined modulo the gauge transformations \cite{FZ,WZ}
\bea
\d_\c H = \c +\bar \c ~, \qquad \bar \nabla^\bd \c =0~,
\eea
with the chiral scalar $\c$ being primary and dimensionless. 
Unlike the $s\geq 1$ case considered above, the vector multiplet field strength $\mathcal{W}_\a = - \frac{1}{4} \bar{\nabla}^2 \nabla_\a H$ 
is gauge invariant, $\d_\c \cW_\a =0$, for an arbitrary supergravity background.


\subsection{$\sU(1)$ duality-invariant models}

It is important to note that the field strength \eqref{3.1} obeys the Bianchi identity\footnote{This differs from the Bianchi identity for the nonlinear super-Weyl tensor \eqref{B.3b} on historical grounds.} \cite{KPR}
\be
\label{3.4}
\nabla^{\b_1}{}_{(\ad_1} \dots \nabla^{\b_s}{}_{\ad_s)} \nabla^{\b_{s+1}} \mathcal{W}_{\a(s) \b(s+1)} 
=-
\nabla_{(\a_1}{}^{\bd_1} \dots \nabla_{\a_s)}{}^{\bd_s} \bar{\nabla}^{\bd_{s+1}} \bar{\mathcal{W}}_{\ad(s) \bd(s+1)}  ~.
\ee
Moreover, the real superfield
\bea
\cB_{\a(s)\ad(s)}:=\nabla^{\b_1}{}_{(\ad_1} \dots \nabla^{\b_s}{}_{\ad_s)} \nabla^{\b_{s+1}} \mathcal{W}_{\a(s) \b(s+1)} 
\eea
proves to be primary,
 \bea
K^{B} \cB_{\a(s) \ad(s)}= 0 ~, \qquad {\mathbb D} \cB_{\a(s) \ad(s)}= (2+s)\cB_{\a(s) \ad(s)}~,
\eea
and may be called the linearised superspin-$(s+\hf)$ Bach tensor. For $s>1$ 
$\cB_{\a(s) \ad(s)}$ is also referred to as  a higher-spin super-Bach tensor. 

We assume that  $S^{(s)}[\mathcal W , \bar{\mathcal W}]$ is consistently defined as a functional of a general chiral superfield $\mathcal{W}_{\a(2s+1)}$ and its conjugate. This allows us to introduce the symmetric spinor
\be
\ri \mathcal{M}_{\a(2s+1)} := 2 \frac{\d S^{(s)}[\mathcal{W},\bar{\mathcal{W}}]}{\d \mathcal{W}^{\a(2s+1)}} ~,
\ee
where the variational derivative is defined by 
\be
\d S^{(s)}[\mathcal{W},\bar{\mathcal{W}}] = \int \rd^4x \rd^2 \q \, \mathcal{E}\, \d \mathcal{W}^{\a(2s+1)} \frac{\d S^{(s)}[\mathcal{W},\bar{\mathcal{W}}]}{\d \mathcal{W}^{\a(2s+1)}}  + \text{c.c.}
\ee
and $\cE$ is the chiral integration measure. 
It follows that $\mathcal{M}_{\a(2s+1)}$ is a primary chiral superfield,
\bea
K^B \cM_{\a(2s+1)} =0~, \qquad \bar \nabla^\bd \mathcal{M}_{\a(2s+1)} =0~,
\qquad 
\mathbb{D} \mathcal{M}_{\a(2s+1)} = \frac 3 2 \mathcal{M}_{\a(2s+1)}~.
\eea
Varying the action with respect to the prepotential $H_{\a(s) \ad(s)}$ yields
the equation of motion
\be
\label{3.8}
\nabla^{\b_1}{}_{(\ad_1} \dots \nabla^{\b_s}{}_{\ad_s)} \nabla^{\b_{s+1}} \mathcal{M}_{\a(s) \b(s+1)} 
= 
- \nabla_{(\a_1}{}^{\bd_1} \dots \nabla_{\a_s)}{}^{\bd_s} \bar{\nabla}^{\bd_{s+1}} \bar{\mathcal{M}}_{\ad(s) \bd(s+1)}  ~.
\ee
Here the real superfield $\nabla^{\b_1}{}_{(\ad_1} \dots \nabla^{\b_s}{}_{\ad_s)} \nabla^{\b_{s+1}} \mathcal{M}_{\a(s) \b(s+1)} $ proves to be primary.

The analysis above indicates that the functional form of \eqref{3.8} coincides with that of \eqref{3.4}. It then follows that the system of equations
\eqref{3.4} and \eqref{3.8} is invariant under the $\sU(1)$ duality rotations:
\be
\label{3.9}
\d_\l \mathcal{W}_{\a(2s+1)} = \l \mathcal{M}_{\a(2s+1)} ~, \quad \d_\l \mathcal{M}_{\a(2s+1)} = - \l \mathcal{W}_{\a(2s+1)} ~,
\ee
where $\l$ is a constant, real parameter. One may then obtain two equivalent expressions for the variation of $S^{(s)}[\mathcal W , \bar{\mathcal W}]$ with respect to \eqref{3.9}
\be
\d_\l S^{(s)}[\mathcal W , \bar{\mathcal W}] = \frac{\ri \l}{4} \int \rd^4x \, \rd^2\q \, \cE \, \Big \{ \cW^2 - \cM^2 \Big \} + \text{c.c.} = -\frac{\ri \l}{2} \int \rd^4x \, \rd^2\q \, \cE \, \cM^2 + \text{c.c.}~,
\label{315}
\ee
as a generalisation of similar derivations in nonlinear $\cN=1$ supersymmetric electrodynamics
\cite{KT1,KT2,KMcC}.\footnote{In eq. \eqref{315} and the remainder of this section we make use of  the notational shorthands $\cW^2 = \cW^{\a(2s+1)} \cW_{\a(2s+1)}$ and $\bar \cW^2 = \bar \cW_{\ad(2s+1)} \bar \cW^{\ad(2s+1)}$.}
This implies the self-duality equation
\be
\label{3.10}
\text{Im} \int \rd^4x \rd^2 \q \, \mathcal{E}\, \Big \{ \mathcal{W}^{\a(2s+1)}
\cW_{\a(2s+1)} + \mathcal{M}^{\a(2s+1)}\cW_{\a(2s+1)} \Big \}  = 0 ~,
\ee
which must hold for a general chiral superfield $\mathcal{W}_{\a(2s+1)}$  and its conjugate. Every solution $S^{(s)}[\mathcal W , \bar{\mathcal W}]$ of the self-duality equation describes a $\sU(1)$ duality-invariant theory.
In the $s=0$ case, the equation \eqref{3.10} was originally derived in \cite{KT1} in Minkowski superspace 
and extended to supergravity in \cite{KMcC}.
The simplest solution of the self-duality equation \eqref{3.10} is the
free $\cN=1$ SCHS model \eqref{D.5}, which was introduced in \cite{KMT} 
for the cases of Minkowski and AdS backgrounds and later was generalised to arbitrary conformally flat backgrounds in \cite{KP}.

The above results allow one to prove, in complete analogy with the non-supersymmetric analysis conducted in section \ref{section2.2}, that the $\sU(1)$ duality-invariant theories are  self-dual under Legendre transformations.


\subsection{Auxiliary variable formulation} \label{section3.2}

As a generalisation of the auxiliary variable formalism sketched in section \ref{section2.3},
here we will develop a reformulation of the $\sU(1)$ duality-invariant systems introduced in the previous subsection. In the $s=0$ case, it will reduce 
to the auxiliary superfield approach for $\sU(1)$ duality-invariant supersymmetric electrodynamics introduced in \cite{K13,ILZ}.

Consider the action functional
\begin{align}
	\label{4.1}
	S^{(s)}[\mathcal{W},\bar{\mathcal W}, \eta, \bar \eta] &= (-1)^{s} \int \rd^4x 
	\rd^2 \q \, \mathcal{E}\, \Big \{ \eta \mathcal{W} - 
	\frac{1}{2} \eta^2- \frac{1}{4}\mathcal{W}^2 \Big \} + \text{c.c.} 
	+ \mathcal{S}^{(s)}_{\text{int}} [\eta , \bar{\eta}] ~.
\end{align}
Here we have introduced the new dimension-$3/2$ multiplet $\eta_{\a(2s+1)}$, which is required to be primary and covariantly chiral, 
\bea
K^B \eta_{\a(2s+1)} =0~, \qquad
\bar{\nabla}_\ad \eta_{\a(2s+1)} = 0~, \qquad {\mathbb D} \eta_{\a(2s+1) } = \frac 32 \eta_{\a (2s+1)}
~.
\eea
By definition, the functional $\mathcal{S}^{(s)}_{\text{int}} [\eta , \bar{\eta}]$,  contains cubic and higher powers of $\eta_{\a(2s+1)}$ and its conjugate.

The equation of motion for $\eta^{\a(2s+1)}$ is 
\be
\label{4.2}
\eta_{\a(2s+1)} = \mathcal{W}_{\a(2s+1)}
+ (-1)^{s} \frac{\d \mathcal{S}^{(s)}_{\text{int}} [\eta , \bar{\eta}]}{\d \eta^{\a(2s+1)}}~.
\ee
Employing perturbation theory, equation \eqref{4.2} allows one to express $\eta_{\a(2s+1)}$ as a functional of $\mathcal{W}_{\a(2s+1)}$ and its conjugate. This means that \eqref{4.1} is dual to a SCHS theory with action
\begin{align}
	\label{4.3}
	S^{(s)}[\mathcal{W},\bar{\mathcal W}] &= \frac{(-1)^s}{4} \int \rd^4x 
	\rd^2 \q \, \mathcal{E}\, \mathcal{W}^2 + \text{c.c.} 
	+ {S}^{(s)}_{\text{int}} [\mathcal{W} , \bar{\mathcal W}] ~.
\end{align}
Thus, \eqref{4.1} and \eqref{4.3} provide two equivalent realisations of the same model.

The power of this formulation is most evident when the self duality equation \eqref{3.10} is applied. A routine computation reveals that this constraint is equivalent to
\be
\label{4.4}
\text{Im} \int \rd^4x \rd^2 \q \, \mathcal{E}\, \eta^{\a(2s+1)} 
\frac{\d \mathcal{S}^{(s)}_{\text{int}} [\eta , \bar{\eta}]}{\d \eta^{\a(2s+1)}} = 0 ~.
\ee
Thus, self-duality of the action \eqref{4.1} is equivalent to the requirement that $\mathcal{S}^{(s)}_{\text{int}}[\eta,\bar{\eta}]$ is invariant under rigid $\sU(1)$ phase transformations
\be
\label{4.5}
\mathcal{S}^{(s)}_{\text{int}} [\re^{\ri \varphi} \eta , \re^{- i \varphi} \bar{\eta}] = \mathcal{S}^{(s)}_{\text{int}} [\eta , \bar{\eta}] ~, \quad \varphi \in \mathbb{R} ~.
\ee

Within the superconformal approach to supergravity-matter dynamical systems \cite{KakuT}, 
every theory of Einstein supergravity coupled to supersymmetric matter is realised as a coupling of the same matter multiplets to conformal supergravity and a superconformal compensator, see, e.g., 
\cite{FGKV}. Truly superconformal theories are independent of any compensator.
Different off-shell formulations for Einstein supergravity correspond to different choices 
of the compensating multiplet. The most famous compensators are: (i) a chiral compensator $S_0$ of the old minimal supergravity; and (ii) a real linear compensator ${\mathbb L}=\overline{\mathbb L}$ of the new minimal supergravity. 
Both $S_0$ and $\mathbb L$ are required to be nowhere vanishing, such that $(S_0)^{-1}$ and 
${\mathbb L}^{-1} $ exist, 
and they satisfy the following superconformal properties:
\begin{subequations}
\bea
K^B S_0&=&0~,\qquad \bar \nabla_\ad S_0 =0~, \qquad {\mathbb D}S_0=S_0~;\\
K^B {\mathbb L}&=&0~, \qquad \bar \nabla^2 {\mathbb L} =0\quad \implies \quad 
{\mathbb D}{\mathbb L}=2{\mathbb L}~.
\eea
\end{subequations}

If we allow the action \eqref{4.3} to depend on a superconformal compensator, either $S_0$ or 
${\mathbb L}$, we can generate families of $\sU(1)$ duality-invariant theories which satisfy the condition 
\eqref{4.5}. For simplicity, unless otherwise stated, here we will restrict our attention to the following interaction
\begin{subequations}\label{467}
\bea
\label{4.6}
\mathcal{S}^{(s)}_{\text{int}} [\eta , \bar{\eta}; \U] &=& \frac{1}{2(2s+2)!} \int \rd^4x \rd^2 \q \rd^2 \bar{\q} \, E \, \frac{ \eta^{2s+2}\bar{\eta}^{2s+2}}{\U^{3s+2}} 
\mathfrak{F}^{(s)} (v, \bar v) ~, 
\eea
where 
\bea
v: = \frac{1}{8} \nabla^2 \big[ {\eta^{2}}{\U^{-2}} \big]~, 
\label{v}
\eea
and the compensator $\U$ 
has one of the two realisations: either  $\U = S_0 \bar S_0$ or $\U={\mathbb L}$.
Applying \eqref{4.5}, we find that our model is $\sU(1)$ duality invariant provided
\bea
\label{4.7}
\mathfrak{F}^{(s)} (v, \bar v) = \mathfrak{F}^{(s)} (v \bar v) ~.
\eea
\end{subequations}


\subsection{Superconformal $\sU(1)$ duality-invariant model}

The $\sU(1)$ duality-invariant model \eqref{467} is superconformal if the action  is independent of $\U$. This uniquely fixes the functional form of $\mathfrak{F}(v \bar v)$ modulo a single real parameter
\be
\mathfrak{F}^{(s)}_{\text{SC}}(v \bar v) = \frac{\kappa}{(v \bar v)^{\frac{1}{4}(3s+2)}} ~, \quad \kappa \in \mathbb{R} ~.
\ee
Employing \eqref{4.2}, we arrive at
\bea
\mathcal{W}_{\a(2s+1)} = \eta_{\a(2s+1)} \bigg \{ 1 &+& \frac{(-1)^s \kappa}{8(2s+2)!} \bar{\nabla}^2 \Big[ (2s+2) \frac{\eta^{2s} \bar{\eta}^{2s+2}}{\U^{3s+2}} \mathfrak{F}^{(s)}_{\text{SC}} \non \\
&&\qquad + \frac{\bar{\eta}^{2s+2}}{4 \U^{3s+2}} \nabla^2 \Big( \eta^{2s+2} \partial_v \mathfrak{F}^{(s)}_{\text{SC}} \Big) \Big] \bigg \} ~,
\eea
which, along with its conjugate, allows us to integrate out the auxiliary variables present in \eqref{4.1}. 
The final result for $s=0$ is the model for superconformal $\sU(1)$ duality-invariant electrodynamics introduced in \cite{K21,BLST2}
\bea
\label{N=1MM}
S^{(0)} [\cW,{\bar \cW}] &=&
\frac{1}{4} \cosh \g \int  \rd^4 x \rd^2 \q  \,\cE \, \cW^2 +{\rm c.c.}
\non \\
&& + \frac{1}{4}\sinh \g   \int \rd^4 x \rd^2 \q \rd^2\bar \q \,E \,
\frac{\cW^2\,{\bar \cW}^2}{\U^2\sqrt{u\bar u} }~,
\eea
where $\g$ is a real coupling constant. 
For $s>0$ the resulting model is
\begin{align} \label{scAction}
	S^{(s>0)}[\mathcal{W},\bar{\mathcal W}] &= \frac{(-1)^s}{4} \int \rd^4x 
	\rd^2 \q \, \mathcal{E}\, \mathcal{W}^2 + \text{c.c.} \non \\
	&\qquad + \frac{\kappa}{2(2s+2)!} \int \rd^4x \rd^2 \q \rd^2 \bar{\q} \, E \, \frac{( \cW^{2}\bar{\cW}^{2})^{s+1}}{\U^{3s+2} (u \bar u)^{\frac 14 (3s+2)}} ~.
\end{align}
In both cases we have made use of the shorthand 
\begin{align}
	u = \frac{1}{8} \nabla^2 \Big[ \frac{ \mathcal{W}^2}{\U^2} \Big] ~.
\end{align}

We note that \eqref{scAction} is invariant under the following rescaling of the conformal compensator
\begin{align}
	\label{CCscaling}
	\U' = \re^{2 \s} \U ~,
\end{align}
which implies that the dependence of \eqref{scAction} on $\U$ is purely superficial. Consequently, the action is superconformal.

It is important to note that, at the component level, the purely bosonic sector of the interaction (the $\k$-term) present in \eqref{scAction} identically vanishes.\footnote{The
numerator in the $\k$-term in \eqref{scAction} contains a product of $4(s+1)$ fermionic superfields. Since the component reduction of this term is computed according to the rule \eqref{3.33b}, it is clear we do not have enough spinor derivatives to convert all 
the fermionic superfields into bosonic ones for $s>0$.
} 
Thus, these actions describe different duality-invariant models than those presented in section \ref{section2.4}.
However, one may construct a supersymmetric duality-invariant model
that contains, for instance, the bosonic theory  \eqref{2.13}
at the component level.

Consider the following supersymmetric duality-invariant model 
\begin{align}
	\label{3.31}
	S^{(s)} [\cW, \bar{\cW}, \eta , \bar{\eta}; \U] &= (-1)^{s} \int \rd^4x 
	\rd^2 \q \, \mathcal{E}\, \Big \{ \eta \mathcal{W} - 
	\frac{1}{2} \eta^2- \frac{1}{4}\mathcal{W}^2 \Big \} + \text{c.c.} \non \\
	&\qquad + \frac{\b}{8} \int \rd^4x \rd^2 \q \rd^2 \bar{\q} \, E \, \frac{ \eta^{2}\bar{\eta}^{2}}{\U^{2} \sqrt{v \bar{v}}} ~.
\end{align}
It does not enjoy invariance under \eqref{CCscaling} and thus \eqref{3.31}  is not superconformal. However,  we are going to demonstrate below that,
 at the component level, this model contains certain conformal duality-invariant 
 actions.
  For this we restrict our attention
to the bosonic sector of \eqref{3.31}. The bosonic field strengths contained in 
$\cW_{\a(2s+1)}$  are:
\begin{subequations}
\begin{align}
	\label{3.33}
	\cC_{\a( 2s+2)} &= \frac{\sqrt 2 \ri }{4} \nabla_{(\a_1 } \cW_{\a_2 \dots  \a_{2s+2}) } |_{\theta = 0}~, \qquad \cC_{\a(2s)} = \frac{1}{4} \sqrt{\frac{2s+1}{s+1}} \nabla^\b \cW_{\b \a(2s) } |_{\theta = 0} ~. 
\end{align}
The bosonic component fields of  $\eta_{\a(2s+1)}$ are:
\begin{align}
	\r_{\a( 2s+2)} &= \frac{\sqrt 2 \ri }{4} \nabla_{(\a_1 } \eta_{\a_2 \dots  \a_{2s+2}) } |_{\theta = 0}~, \qquad \r_{\a(2s)} = \frac{1}{4} \sqrt{\frac{2s+1}{s+1}} \nabla^\b \eta_{\b \a(2s) } |_{\theta = 0} ~.
\end{align}
\end{subequations}
Action \eqref{3.31} can be reduced to components using 
 the standard reduction rules
 \begin{subequations}
\bea
\int \rd^4 x \rd^2 \q\, {\cE} {\cL}_c   &=&-{1\over4}\int \rd^4 x    \,e 
\nabla^2 {\cL}_c |_{\q =0} ~, \qquad \bar \nabla^\ad {\cL}_c =0~, 
 \\
\int \rd^4x \rd^2 \q \rd^2 \bar \q \, {E} \, {\cL} &=&{1\over16}\int \rd^4x \, e\,
\nabla^2 \, \bar \nabla^2 {\cL}|_{\q =0} ~.
\label{3.33b}
\eea
\end{subequations}
Since we are interested only in the bosonic sector, it suffice to approximate
\bea
v|_{\q=0} \approx \frac{2}{(\U|)^2} \Big\{ \r^{\a(2s+2)} \r_{\a(2s+2)} 
- \r^{\a(2s)} \r_{\a(2s)}  \Big\}~,\qquad \U| \equiv \U|_{\q=0}~.
\eea 
Further, it may be checked that  the interaction (the $\b$-term)  in \eqref{3.31} does not contain contributions  which are (i) linear in $\r_{\a(2s)}$ and its conjugate; and (ii)  linear in $\r_{\a(2s+2)}$ and its conjugate.
Therefore, if 
we switch off the spin-$s$ field, 
\bea
\label{switchoffGF}
\cC_{\a(2s)} = 0~,
\eea
then 
\bea
\r_{\a(2s)} = 0
\label{switchoffRF}
\eea
is a solution of the corresponding  equation of motion.
Under the conditions \eqref{switchoffGF} and \eqref{switchoffRF}, 
the resulting bosonic  action proves to coincide with 
$S^{(s+1)}[\mathcal{C},\bar{\mathcal C}, \r, \bar \r]  $
obtained from \eqref{AuxAction} by replacing $s$ with $s+1$.
We emphasise that the compensator $\U$ does not contribute to this action, and therefore $S^{(s+1)}[\mathcal{C},\bar{\mathcal C}, \r, \bar \r] $
is locally conformal.\footnote{As demonstrated in appendix \ref{appendixNew}, the action \eqref{AuxAction} leads to \eqref{2.13} upon eliminating the auxiliary field $\r_{\a(2s)}$ and its conjugate.} 
Instead of considering the branch  \eqref{switchoffGF},
we may switch off the spin-$(s+1) $ field, $\cC_{\a(2s+2)}=0$.
  Then it follows that $\r_{\a(2s+2)}=0$ is a solution of the corresponding equation of motion, and the resulting bosonic  action proves to coincide with 
the conformal duality-invariant action $S^{(s)}[\mathcal{C},\bar{\mathcal C}, \r, \bar \r]  $, eq. \eqref{AuxAction}.

To conclude this section, we derive a supersymmetric extension of the cubic interaction present in \eqref{2.35}. For this purpose, we will fix $s=1$ in \eqref{4.1}. The relevant interaction is constructed in terms of the two scalar primary descendants of $\eta_{\a(3)}$:
\begin{align}
	\Xi = \ri \eta^{\a(2) \b} \eta^{\a(2)}{}_\b \nabla_{(\a_1} \eta_{\a_2 \a_3 \a_4)} ~, \qquad
	w = 4 \nabla^2 \Big[ \frac{ \Xi }{\U^3} \Big] ~,
\end{align}
and the conformal compensator $\U$. Using these, we construct the following action
\begin{align}
	\label{3.35}
	S^{(1)} [\cW, \bar{\cW}, \eta , \bar{\eta}; \U] &= - \int \rd^4x 
	\rd^2 \q \, \mathcal{E}\, \Big \{ \eta \mathcal{W} - 
	\frac{1}{2} \eta^2- \frac{1}{4}\mathcal{W}^2 \Big \} + \text{c.c.} \non \\
	&\qquad + 32 \kappa \int \rd^4x \rd^2 \q \rd^2 \bar{\q} \, E \, \frac{ \Xi \bar{\Xi} }{\U^{4} (w \bar{w})^{\frac 23}} ~.
\end{align}
We note that it is not superconformal as it lacks invariance under \eqref{CCscaling}, though we will show that, at the component level, it contains the cubic interaction present in \eqref{2.35}. To demonstrate this, we restrict our attention to the bosonic sector of \eqref{3.35}, which allows us to approximate
\begin{align}
\label{w|}
\quad w |_{\q = 0} \approx - \frac{\sqrt 2}{4 (\U|)^3}\Big( \r^{\a(2)}{}_{\b(2)} \r^{\b(2)}{}_{\g(2)} \r^{\g(2)}{}_{\a(2)} - \frac{9}{32} \r^{\a(2)} \r^{\a(2)} \r_{\a(4)} \Big) ~.
\end{align}
Further, the interaction (the $\k$-term) present in \eqref{3.35} contains no terms linear in either $\r_{\a(4)}$ or $\r_{\a(2)}$. Hence, if we switch off the spin-$1$ field, $\cC_{\a(2)} = 0$, then $\r_{\a(2)} = 0$ solves the corresponding equations of motion. Under these conditions, the resulting bosonic action takes the form
\begin{align}
	S^{(2)}[\mathcal{C},\bar{\mathcal C}, \r, \bar \r] &= \int \rd^4x 
	\, e \, \Big \{ 2 \r \mathcal{C} - 
	\r^2- \frac{1}{2}\mathcal{C}^2 \Big \} + \text{c.c.} + \kappa \int \rd^4x 
	\, e \, \big( \r^3 \bar \r^3\big)^{\frac 13} ~.
\end{align}
In particular, we note that it is locally conformal as it is independent of the compensator. Thus, \eqref{3.35} is a supersymmetric extension of the cubic interaction of \eqref{2.35}. In closing, we emphasise that if one instead attempts to consider the branch $\cC_{\a(4)} = 0$ and $\cC_{\a(2)} \neq 0$, it is not legal to set $\r_{\a(4)} = 0$. Such a choice would imply $w|_{\q=0} \approx 0$, making the corresponding component action of \eqref{3.35} undefined.


\section{$\mathcal{N} = 2$ duality-invariant SCHS models} \label{section4}

This section is devoted to a brief study of duality-invariant  models for $\N=2$ conformal 
superspin-$(s+1)$ gauge multiplets, with $s\geq 0$, in  conformally flat backgrounds.\footnote{See, e.g., 
\cite{RS,GGRS} for the notions  of $\cN=2$ superspin and superisospin. For all superconformal multiplets considered in this section, their superisospin is equal to zero. } 
Our approach will generalise the general theory of $\sU(1)$ duality-invariant models for $\cN=2$ 
supersymmetric nonlinear electrodynamics (superspin-$0$) developed in \cite{KT1,KT2,K12,K13}.
The essential information about $\cN=2$ conformal superspace, which is used throughout this section, 
is collected in appendix \ref{appendixC}.

We start with a review of superconformal higher-spin (SCHS) gauge multiplets \cite{KR21}.
Given a non-negative integer $s$, the conformal superspin-$(s+1)$ gauge multiplet is described by a real unconstrained prepotential ${H}_{\a(s) \ad(s)}$, which is a primary superfield,
$K^B {H}_{\a(s) \ad(s)}=0$,  defined modulo gauge transformations of the form
\begin{subequations}
\label{GT}
\bea
\label{N=2gt}
&s>0:& \qquad \d_\z H_{\a(s) \ad(s)} = \nabla_{(\a_1}^i \bar{\z}_{\a_2 \dots \a_s) \ad(s)i} + \bar{\nabla}_{(\ad_1}^i \z_{\a(s) \ad_2 \dots \ad_s) i}~, \\
&s=0:& \qquad\d_\z H = \nabla^{ij} \bar{\z}_{ij} + \bar{\nabla}^{ij} \z_{ij} ~,
\eea
\end{subequations}
with the gauge parameters $\z_{\a(s) \ad(s-1) i} $ and $\z_{ij}$ being primary and complex unconstrained. Here we have defined the second-order operators 
\bea
\bar{\nabla}^{ij} := \bar{\nabla}_{\ad}^{(i} \bar{\nabla}^{\ad j)} 
~,\qquad
{\nabla}^{ij} := {\nabla}^{\a(i} \bar{\nabla}_\a^{ j)} ~.
\eea
The gauge transformation laws \eqref{GT} are superconformal provided 
\bea
\mathbb{D} H_{\a(s) \ad(s)} = - (s+2) H_{\a(s) \ad(s)} ~.
\eea

Associated with the  gauge prepotential ${H}_{\a(s) \ad(s)}$ is the chiral descendant \cite{KR21}
\be
\label{5.1}
\bm{\mathcal{W}}_{\a(2s+2)} =  \bar{\nabla}^4 \nabla_{(\a_1}{}^{\bd_1} \dots \nabla_{\a_s}{}^{\bd_s} \nabla_{\a_{s+1} \a_{s+2}} {H}_{\a_{s+3} \dots \a_{2s+2}) \bd(s)} ~,
\qquad \bar \nabla^\bd_j \bm{\mathcal{W}}_{\a(2s+2)} = 0~,
\ee
where we have introduced 
the chiral projection operator 
\bea
\bar{\nabla}^4  := \frac{1}{48} \bar{\nabla}^{ij} \bar{\nabla}_{ij} 
\eea
and the second-order operators 
\begin{align}
\nabla_{\a\b} := \nabla_{(\a}^k \nabla_{\b) k} \ , \qquad
 \bar{\nabla}^{\ad\bd} := \bar\nabla^{(\ad}_k \bar\nabla^{\bd) k} \ .
\end{align}
It can be shown that $\bm{\mathcal{W}}_{\a(2s+2)} $ is primary in an arbitrary supergravity background,
\bea
K^B \bm{\mathcal{W}}_{\a(2s+2)} =0~, \qquad
 \mathbb{D} \bm{\mathcal{W}}_{\a(2s+2)} = \bm{\mathcal{W}}_{\a(2s+2)}~.
 \eea
 However, the gauge transformations \eqref{GT} leave $\bm{\mathcal{W}}_{\a(2s+2)}$ invariant, $\d_\z \bm{\mathcal{W}}_{\a(2s+2)} = 0$, only if the background curved superspace is conformally flat,
\be
W_{\a(2)} = 0 \quad \implies \quad \d_\z \bm{\mathcal{W}}_{\a(2s+2)} = 0~,
\ee
where $W_{\a(2)}$ is the background super-Weyl tensor, see appendix \ref{appendixC}.
Throughout this section, we will restrict ourselves to such a geometry.

We should point out that the gauge prepotential $H$ describes the linearised $\cN=2$ conformal supergravity multiplet, 
and its field strength $\bm{\mathcal{W}}_{\a(2)} $ is the linearised super-Weyl tensor. For $s>0$  we will refer to
the fields strengths $\bm{\mathcal{W}}_{\a(2s+2)}$ as the linearised  higher-spin super-Weyl tensors.

The chiral field strength defined by \eqref{5.1} carries at least two spinor indices. 
A chiral scalar field strength $\bm{\cW}$ corresponds to  the massless $\N=2$ vector multiplet \cite{GSW}. It can be described in terms of the curved superspace analogue
of Mezincescu's prepotential \cite{Mezincescu} (see also \cite{HST}),
 $V_{ij}=V_{ji}$,
which is an unconstrained real SU(2) triplet, 
$\overline{V_{ij}} = V^{ij}= \ve^{ik}\ve^{jl}V_{kl}$. 
The expression for $\bm{\cW}$ in terms of $V_{ij}$ 
was found in \cite{ButterK} to be
\bea
\bm{\cW} = \frac{1}{4}\bar\nabla^4 \nabla^{ij} V_{ij}~, \quad \bar{\nabla}^{\bd}_j \bm{\cW} = 0 ~, \quad V_{ij} = V_{ji}~.
\eea
The field strength $\bm{\cW}$ defines a primary reduced chiral superfield of dimension $+1$, 
\bea
\nabla^{ij} \bm{\cW} = \bar \nabla^{ij} \bar{\bm{\cW}}~, \quad \mathbb{D} \bm{\cW} = \bm{\cW} ~, \quad K_A \bm{\cW} = 0~,
\eea
and is invariant under gauge transformations of the form \cite{ButterK2}
\bea
\delta_\z V^{ij} &= \nabla^{\alpha}{}_k \z_\alpha{}^{kij}
+ \bar\nabla_{\dalpha}{}_k \bar\z^\dalpha{}^{kij}, \qquad
\z_\alpha{}^{kij} = \z_\alpha{}^{(kij)}~,
\label{pre-gauge}
\eea
with $ \z_\alpha{}^{kij} $ being primary and unconstrained modulo the algebraic condition given. 
In the flat-superspace limit, the gauge transformation law \eqref{pre-gauge} reduces to that given in 
\cite{Mezincescu,HST}. It is important to emphasise that $\bm{\cW}$ is invariant under the gauge transformations \eqref{pre-gauge} in an arbitrary supergravity background.

The formalism we introduce below for nonlinear SCHS models generalises the existing framework for $\sU(1)$ duality-invariant theories of the vector multiplet. The latter have been studied extensively in  \cite{KT1,KT2,K13,K21}, thus we omit this special case from our considerations.


\subsection{$\sU(1)$ duality-invariant models}

We consider a dynamical system describing the propagation of 
the superconformal gauge  multiplet ${H}_{\a(s) \ad(s)}$, $s\geq 0$, 
  in a conformally flat superspace. We require its action functional to take the form $S^{(s)}[\bm{\mathcal{W}}, \bar{\bm{\mathcal W}}]$, where the gauge-invariant chiral field strength 
 $\bm{\mathcal{W}}_{\a(2s+2)}$ is given by eq. \eqref{5.1}.

It is important to note that the field strength \eqref{5.1} obeys the Bianchi identity
\be
\nabla^{\b_1}{}_{(\ad_1} \dots \nabla^{\b_s}{}_{\ad_s)} \nabla^{\b_{s+1} \b_{s+2}} \bm{\mathcal{W}}_{\a(s) \b(s+2)} 
=
\nabla_{(\a_1}{}^{\bd_1} \dots \nabla_{\a_s)}{}^{\bd_s} \bar{\nabla}^{\bd_{s+1} \bd_{s+2}} \bar{\bm{\mathcal{W}}}_{\ad(s) \bd(s+2)}  ~.
\label{4.12}
\ee
We then assume  that $S^{(s)}[\bm{\mathcal W} , \bar{\bm{\mathcal W}}]$ can be defined as a functional of a general chiral superfield $\bm{\mathcal{W}}_{\a(2s+2)}$ and its conjugate. This allows us to introduce the symmetric spinor
\be
\ri \bm{\mathcal{M}}_{\a(2s+2)} := 2 \frac{\d S^{(s)}[\bm{\mathcal{W}},\bar{\bm{\mathcal{W}}}]}{\d \bm{\mathcal{W}}^{\a(2s+2)}} ~,
\ee
which defines an arbitrary variation of the action
\be
\d S^{(s)}[\bm{\mathcal{W}},\bar{\bm{\mathcal{W}}}] = \int \rd^4x \rd^2 \q \, 
\bm{\mathcal{E}}\, \d \bm{\mathcal{W}}^{\a(2s+2)} \frac{\d S^{(s)}[\bm{\mathcal{W}},\bar{\bm{\mathcal{W}}}]}{\d \bm{\mathcal{W}}^{\a(2s+2)}}  + \text{c.c.}~,
\ee
where $\bm \cE$ is the $\cN=2$ chiral integration measure.\footnote{The chiral and full superspace integrals are related to each according to eq. \eqref{C.9}.}
It follows that $\bm{\mathcal{M}}_{\a(2s+2)}$ is a primary chiral superfield, 
\bea
K^B \bm{\mathcal{M}}_{\a(2s+2)}=0~, \qquad \bar \nabla^\bd_j \bm{\mathcal{M}}_{\a(2s+2)} =0~, 
\qquad {\mathbb D} \bm{\mathcal{M}}_{\a(2s+2)}= \bm{\mathcal{M}}_{\a(2s+2)}~.
\eea
Varying the action with respect to the prepotential $H_{\a(s) \ad(s)}$ yields the equation of motion
\be
\nabla^{\b_1}{}_{(\ad_1} \dots \nabla^{\b_s}{}_{\ad_s)} \nabla^{\b_{s+1} \b_{s+2}} \bm{\mathcal{M}}_{\a(s) \b(s+2)} 
= 
\nabla_{(\a_1}{}^{\bd_1} \dots \nabla_{\a_s)}{}^{\bd_s} \bar{\nabla}^{\bd_{s+1}\bd_{s+2}} \bar{\bm{\mathcal{M}}}_{\ad(s) \bd(s+2)}  ~.
\label{4.16}
\ee

This analysis above indicates that the functional form of \eqref{4.16} coincides with that of \eqref{4.12}. 
As a result, the system of equations \eqref{4.12} and \eqref{4.16} is invariant under $\sU(1)$ duality rotations:
\be
\label{N=2dualityRot}
\d_\l \bm{\mathcal{W}}_{\a(2s+2)} = \l \bm{\mathcal{M}}_{\a(2s+2)} ~, \quad \d_\l \bm{\mathcal{M}}_{\a(2s+2)} = - \l \bm{\mathcal{W}}_{\a(2s+2)} ~,
\ee
where $\l$ is a constant, real parameter. One may then obtain two equivalent expressions for the variation of $S^{(s)}[\bm{\cW} , \bar{\bm{\cW}}]$ with respect to \eqref{N=2dualityRot}
\be
\d_\l S^{(s)}[\bm{\cW} , \bar{\bm{\cW}}] = \frac{\ri \l}{4} \int \rd^4x \, \rd^4\q \, \bm{\cE} \, \Big \{ \bm{\cW}^2 - \bm{\cM}^2 \Big \} + \text{c.c.} 
= -\frac{\ri \l}{2} \int \rd^4x \, \rd^4\q \, \bm{\cE} \, \bm{\cM}^2 + \text{c.c.}~,
\ee
as a generalisation of similar derivations in nonlinear $\cN=2$ supersymmetric electrodynamics
\cite{KT1,KT2,K12}.
This implies the self-duality equation
\be
\label{4.10}
\text{Im} \int \rd^4x \rd^4 \q \, \bm{\mathcal{E}}\, \Big \{ \bm{\mathcal{W}}^{\a(2s+2)}
\bm{\cW}_{\a(2s+2)}
 + \bm{\mathcal{M}}^{\a(2s+2)} \cM_{\a(2s+2)} \Big \}  = 0 ~,
\ee
which must hold for a general chiral superfield $\bm{\cW}_{\a(2s+2)}$  and its conjugate. Every solution $S^{(s)}[\bm{\cW} , \bar{\bm{\cW}}]$ of the self-duality equation describes a $\sU(1)$ duality-invariant theory.
Equation \eqref{4.10} allows one to prove, in complete analogy with the non-supersymmetric analysis conducted in section \eqref{section2.2}, that the $\sU(1)$ duality-invariant theories  are self-dual under Legendre transformations.
The simplest solution of the self-duality equation \eqref{4.10} is the
free $\cN=2$ SCHS model \eqref{D.8}  proposed in \cite{KR21}, including the action for linearised conformal supergravity ($s=0$).

\subsection{Auxiliary variable formulation} \label{section4.2}

This subsection is aimed at generalising the auxiliary variable formalism employed in section \ref{section3.2} to construct self-dual $\N=1$ SCHS models to an $\N=2$ setting.\footnote{Such a setup was recently utilised in \cite{K21} to construct $\N=2$ superconformal duality-invariant models for an Abelian vector multiplet in a conformal supergravity background.} To this end, we consider the action functional
\begin{align}
	\label{N=2Aux}
	S^{(s)} [\bm{\mathcal{W}},\bar{\bm{\mathcal W}}, \bm{\eta}, \bar{\bm{\eta}}] &= (-1)^{s} \int \rd^4x 
	\rd^4 \q \, \bm{\mathcal{E}}\, \Big \{ \bm{\eta} \bm{\mathcal{W}} - 
	\frac{1}{2} \bm{\eta}^2- \frac{1}{4}\bm{\mathcal{W}}^2 \Big \} + \text{c.c.} + \mathcal{S}^{(s)}_{\text{int}} [\bm{\eta} , \bar{\bm{\eta}}] ~,
\end{align}
where we have introduced the auxiliary dimension-$1$ multiplet $\bm{\eta}_{\a(2s+2)}$, which is primary and covariantly chiral, 
\bea
K^B \bm{\eta}_{\a(2s+2)} = 0~,\qquad
\bar{\nabla}^\bd_j \bm{\eta}_{\a(2s+2)} = 0~, \qquad {\mathbb D} \bm{\eta}_{\a(2s+2) } = \bm{\eta}_{\a (2s+2)}
~.
\eea
By construction, the self-interaction $\mathcal{S}^{(s)}_{\text{int}} [\bm{\eta} , \bar{\bm{\eta}}]$  contains cubic and higher powers of $\bm{\eta}_{\a(2s+2)}$ and its conjugate. Varying \eqref{N=2Aux} with respect to this superfield yields 
\be
\label{5.2}
\bm{\eta}_{\a(2s+2)} = \bm{\mathcal{W}}_{\a(2s+2)}
+ (-1)^{s} \frac{\d \mathcal{S}^{(s)}_{\text{int}} [\bm{\eta} , \bar{\bm{\eta}}]}{\d \bm{\eta}^{\a(2s+2)}}~.
\ee
Employing perturbation theory, equation \eqref{5.2} allows one to express $\bm{\eta}_{\a(2s+2)}$ as a functional of $\bm{\mathcal{W}}_{\a(2s+1)}$ and its conjugate. This means that \eqref{N=2Aux} is equivalent to a SCHS theory with action
\begin{align}
	\label{5.3}
	S^{(s)}[\bm{\mathcal{W}},\bar{\bm{\mathcal W}}] &= \frac{(-1)^s}{4} \int \rd^4x 
	\rd^4 \q \, \bm{\mathcal{E}}\, \bm{\mathcal{W}}^2 + \text{c.c.} 
	+ {S}^{(s)}_{\text{int}} [\bm{\mathcal{W}} , \bar{\bm{\mathcal W}}] ~.
\end{align}
Thus, \eqref{N=2Aux} and \eqref{5.3} provide two equivalent realisations of the same model.

We now analyse the self duality equation \eqref{4.10}. It is easily verified that  this constraint is equivalent to
\be
\label{5.4}
\text{Im} \int \rd^4x \rd^4 \q \, \bm{\mathcal{E}}\, \bm{\eta}^{\a(2s+2)} 
\frac{\d \mathcal{S}^{(s)}_{\text{int}} 
[\bm{\eta} , \bar{\bm{\eta}}]}{\d \bm{\eta}^{\a(2s+2)}} = 0 ~.
\ee
Thus, the $\sU(1)$ duality invariance of the model \eqref{N=2Aux} is equivalent to the requirement that $\mathcal{S}^{(s)}_{\text{int}}[\bm{\eta},\bar{\bm{\eta}}]$ is invariant under rigid $\sU(1)$ phase transformations
\be
\label{5.5}
\mathcal{S}^{(s)}_{\text{int}} [\re^{\ri \varphi} \bm{\eta} , \re^{- i \varphi} \bar{\eta}] = \mathcal{S}^{(s)}_{\text{int}} [\bm{\eta} , \bar{\bm{\eta}}] ~, \quad \varphi \in \mathbb{R} ~.
\ee

If we allow the action \eqref{N=2Aux} to depend on a superconformal compensator, it is 
trivial to construct interactions satisfying the condition \eqref{5.5}, for instance 
\bea
\mathcal{S}^{(s)}_{\text{int}} [\bm{\eta} , \bar{\bm{\eta}} ; {\bm \cW}_0 , \bar {\bm \cW}_0] =
 \int \rd^4x \rd^4 \q \rd^4 \bar \q \, \bm{E}\, {\mathfrak F} \Big( 
 \frac{ \bm{\eta}^{\a(2s+2)} \bm{\eta}_{\a(2s+2) } \bar{\bm{\eta}}_{\ad(2s+2)} 
 \bar{\bm{\eta}}^{\ad(2s+2)} 
 }
 { ({\bm \cW}_0 \bar {\bm \cW}_0)^2 } \Big)~.
 \eea
Here $ {\mathfrak F} (x)$ is a real analytic function of a real variable, and ${\bm \cW}_0$ is the chiral field strength of a compensating vector multiplet. 

It is not difficult to construct explicit examples of $\cN=2$ superconformal 
and $\sU(1)$ duality-invariant higher-spin theories, as a generalisation of the $\cN=2$ vector multiplet models presented in \cite{K21}. For simplicity, here we restrict our attention to $\cN=2$ Minkowski superspace. Let us consider the self-interaction
\bea
\mathcal{S}^{(s)}_{\text{int}} [\bm{\eta} , \bar{\bm{\eta}}] =
c \int \rd^4x \rd^4 \q \rd^4 \bar \q \,  \ln 
  \bm{\eta}^2
  \ln   \bar{\bm{\eta}}^2~,
 \eea
 with $c$ a coupling constant. It is clearly  invariant under rigid $\sU(1)$ phase transformations, eq. \eqref{5.5}, and therefore it generates a $\sU(1)$ duality-invariant model. It is also $\cN=2$ superconformal, which may be checked using the properties
 of $\cN=2$ superconformal transformations \cite{KT-superconformal,BKT}. The above functional is well defined provided the auxiliary variables $\bm{\eta}_{\a(2s+2)}$
 are chosen to belong to the open domain
 $  \bm{\eta}^2:= \bm{\eta}^{\a(2s+2)} \bm{\eta}_{\a(2s+2) } \neq 0$.
A similar condition should be imposed in the case of non-supersymmetric model 
\eqref{confgravitonNLAction}.


\subsection{On $\cN=2$ super ModMax theory}

So far, an $\cN=2$ superconformal extension of the ModMax theory has not been constructed. Perhaps it does not exist (see the discussion in \cite{K21}), the main reason being the fact that the full $\cN=2$ superspace measure is dimensionless.
However, if the action functional is allowed to depend on compensators, 
an $\cN=2$ supersymmetric extension of the ModMax theory can be introduced. Specifically, let us consider the following $\cN=2$ vector multiplet model\footnote{Our normalisation of the quadratic part of the vector multiplet action follows \cite{K21}}:
\bea
\label{N=2MM}
S[ \bm{\cW}, \bar{\bm{\cW}}, \bm{\eta}, \bar{\bm{\eta}}; {\bm \cW}_0, \bar{{\bm \cW}}_0]&=& \hf \int \rd^4 x \rd^4 \q  \,\bm{\cE}
\, \Big\{ \bm{\eta} \bm{\cW} -\hf \bm{\eta}^2 - \frac{1}{4} \bm{\cW}^2\Big\} 
+{\rm c.c.} \non \\
&& +\mathcal{S}_{\text{int}} [\bm{\eta} , \bar{\bm{\eta}} ; {\bm \cW}_0, \bar{{\bm \cW}}_0 ] \non \\
\mathcal{S}_{\text{int}} [\bm{\eta} , \bar{\bm{\eta}} ; {\bm \cW}_0, \bar{{\bm \cW}}_0 ] &=& \k
 \int \rd^4x \rd^4 \q \rd^4 \bar \q \, \bm{E}\, \frac{ \bm{\eta}^2 \bar{\bm{\eta}}^2 } 
{
 {\bm \cW}_0 \bar {\bm \cW}_0 \sqrt{
 \nabla^4  \Big[ \frac{\bm{\eta} }{{\bm \cW}_0}\Big]^2
\bar  \nabla^4  \Big[ \frac{\bar{\bm{\eta} }}{\bar{{\bm \cW}}_0}\Big]^2
 }}~, ~~~
\eea
where $\k \in \mathbb{R}$ is the coupling constant. We are going to show that this model incorporates  $\cN=1$  superconformal ModMax theory, eq.  \eqref{N=1MM},
in the sense described below.

The field strength $ \bm{\cW}$ of the $\cN=2$ vector multiplet 
contains two independent  $\cN=1$ chiral components, $\F$ and $\cW_\a$,
defined by 
\bea
\bm{\cW} |_{\q_\2 =0}  = \sqrt{2}\, \F~, \qquad  
\nabla_\a^{\underline{2}}\, \bm{\cW} |_{\q_\2 =0} = 2{\rm i}\, \cW_\a ~ \implies ~(\nabla^{\underline{2}})^2 \bm{\cW}|_{\q_\2 =0} 
=\sqrt{2} \, \bar{\nabla}^2\bar\F~.
\eea
Here $\cW_\a$ is the chiral field strength of the $\cN=1$ vector multiplet, 
$\nabla^\a \cW_\a = \bar \nabla_\ad \bar \cW^\ad$. 
The auxiliary $\cN=2$ chiral superfield $ \bm{\eta}$ contains three independent 
$\cN=1$ chiral components, $\c$, $\eta_\a$ and $\J$, defined by 
\bea
\bm{\eta} |_{\q_\2 =0}  = \sqrt{2}\, \c~, \qquad  
\nabla_\a^{\underline{2}}\, \bm{\eta} |_{\q_\2 =0} = 2{\rm i}\, \eta_\a~,\qquad 
-\frac 14 (\nabla^{\underline{2}})^2 \bm{\eta}|_{\q_\2 =0} 
=\J~.
\eea
Using the standard $\cN=2 \to \cN=1$ reduction rules
\bea
\int \rd^4 x \rd^4 \q\, \bm{\cE} \bm{\cL}_c   &=&-{1\over4}\int \rd^4 x \rd^2 \q   \,\cE 
(\nabla^{\underline{2}})^2 \bm{\cL}_c |_{\q_\2 =0} ~, \qquad \bar \nabla^\ad_i \bm{\cL}_c =0~, 
\non \\
\int \rd^4x \rd^4 \q \rd^4 \bar \q \, \bm{E} \, \bm{\cL} &=&{1\over16}\int \rd^4x \rd^2 \q \rd^2 \bar \q \, E\,
(\nabla^{\underline{2}})^2 \, (\bar \nabla_{\underline{2}})^2 \bm{\cL}|_{\q_\2 =0} ~,
\eea
we may reduce the action \eqref{N=2MM} to $\N=1$ superspace. Here we restrict our attention to the $\cN=1$ vector multiplet and switch off the $\cN=1$ scalar multiplet, 
$
\Phi = 0$.
Then it may be shown that setting 
$\c =0$ and $ \J =0$
gives a solution of the equations of motion for these auxiliary superfields.\footnote{This follows form the fact that  the $\cN=1$ superspace reduction of \eqref{N=2MM} 
does not have terms linear in $\c$, $\J$ and their conjugates provided $\F=0$.}
The resulting $\N=1$ superspace action is
\begin{align}
	S[\mathcal{W},\bar{\mathcal W}, \eta, \bar \eta] &= \int \rd^4x 
	\rd^2 \q \, \mathcal{E}\, \Big \{ \eta \mathcal{W} - 
	\frac{1}{2} \eta^2- \frac{1}{4}\mathcal{W}^2 \Big \} + \text{c.c.} \non \\
	& \quad + \k \int \rd^4 x \rd^2 \q \rd^2\bar \q \,E \,
	\frac{\eta^2\,{\bar \eta}^2}{ \U^2 \sqrt{v \bar{v}} } ~.
	\label{4.32}
\end{align}
Here $v$ is defined as in \eqref{v}, and 
we utilise the realisation $\U = S_0 \bar{S}_0$ for the compensator, with  $\bar{\bm{\cW}}_0| = S_0$. 
The action \eqref{4.32} was proposed in \cite{K21} to describe 
$\cN=1$ superconformal $\sU(1)$ duality-invariant electrodynamics. 
Upon eliminating the auxiliary chiral spinor $\eta_\a$ and its conjugate, this model was shown in \cite{K21} to reduce to $\cN=1$  superconformal ModMax theory, eq.  \eqref{N=1MM}, introduced in \cite{BLST2}.


\section{Conclusion} 
\label{section5}

In this paper we have generalised the general theory of $\sU(1)$ duality-invariant nonlinear electrodynamics \cite{GR2,GZ2,GZ3,IZ_N3,IZ1,IZ2} and its $\cN=1$ and $\cN=2$ supersymmetric extensions \cite{KT1,KT2,KMcC,KMcC2,K12,K13,ILZ} to the cases of bosonic conformal spin-$s$ gauge fields, $s\geq 2$, and their $\cN=1$ and $\cN=2$ superconformal cousins. These self-dual higher-spin theories share several important features of duality-invariant electrodynamics, such as self-duality under Legendre transformations. The crucial difference between the $\sU(1)$ duality-invariant models for electrodynamics and their higher-spin counterparts is that the former are consistently defined on arbitrary gravitational backgrounds, while the latter are formulated in a conformally flat space. 

In the non-supersymmetric case, we presented several families of self-dual models. The simplest of these is \eqref{2.13}, which is a higher spin generalisation of ModMax electrodynamics \cite{BLST}. We also introduced a two-parameter family of models for the conformal graviton \eqref{2.35} and provided insights on the construction of new families of multi-parameter duality invariant models for conformal higher spin fields. Finally, in the supersymmetric case, we generalised the constructon of $\N=1$ superconformal duality invariant electrodynamics \cite{BLST2,K21} to higher spins \eqref{scAction}. A distinguishing feature of this family of models is that, at the component level, their purely bosonic sectors vanish, which may be verified via a component reduction.

It should be emphasised that the literature on duality is huge and, unfortunately, it is not possible to mention all publications here. In this paper we were interested in duality as a continuous symmetry of the equations of motion.
 There exists a different approach to duality as a manifest symmetry of the action. The latter was advocated in  
\cite{Deser:1976iy,Deser:1981fr,Henneaux:2004jw,Deser:2004xt,Bunster:2011qp,Bunster:2012jp,BHH}. Another important approach was inspired by early examples 
of dual field theories \cite{OP,KalbR,CS,Curtright1,FreedmanT,Curtright2,FT85} and resulted in various dual formulations for massless, massive and partially massless higher-spin field theories, see  \cite{Hull,BekaertB,BCH,MV,Zinoviev,Hinterbichler,BCC} and references therein.
This approach corresponds to discrete dualities.

There may be several generalisations of our results that are modelled on similar properties of duality-invariant theories of spin-1 gauge fields, see \cite{AFZ} for a review. In particular, we recall that, given a $\sU(1)$ duality-invariant model for nonlinear electrodynamics, its  compact duality group $\sU(1)$
can be enhanced to the non-compact $\sSL(2,{\mathbb R})$ group by coupling the electromagnetic field to the dilaton and axion \cite{GR2,GZ2,GZ3}. It may be shown that this construction naturally extends to the higher-spin case.
We should point out that a generalisation of the method of \cite{IZ_N3,IZ1,IZ2} 
to the case of $\sSL(2,{\mathbb R})$ duality
was given by Ivanov, Lechtenfeld and Zupnik \cite{ILZ2}.

Fundamental to our analysis in this paper has been the formalism of conformal (super) space, which trivialises calculations in backgrounds with vanishing (super) Weyl tensor. For the purpose of applications, however, it is often useful to work with Lorentz covariant derivatives as opposed to their conformally covariant counterparts. The process of translating results expressed in terms of the latter to the former is known as degauging. The general prescription for such an analysis is well-known (see e.g. \cite{KP,ButterN=1,ButterN=2}), however such calculations are often highly non-trivial on generic curved backgrounds. If the background geometry is restricted by turning off several components of the curvature, the resulting computations are greatly simplified. To this end, we now provide a dictionary to translate our main results to the AdS (super)space.\footnote{The conformal flatness of  $\mathcal{N}$-extended  AdS superspace in four dimensions was established in \cite{BILS} and further elaborated in \cite{KT-M-ads}.}

We begin with the non-supersymmetric story. The geometry of $\text{AdS}_4$ is encoded within the Lorentz covariant derivatives $\cD_a$, which obey the algebra
\be
\big[ \cD_\aa , \cD_\bb \big] = - 2 \mu \bar{\mu} \big( \ve_{\ad \bd} M_{\a \b} + \ve_{\a \b} \bar{M}_{\ad \bd} \big) ~,
\ee
where $\mu$ is a nonzero complex constant with the dimension of mass encoding the curvature of the spacetime and $\mu \bar{\mu}$ is proportional to the scalar curvature. It may be shown that in this geometry, equation \eqref{2.1} takes the form
\be
\mathcal{C}_{\a(2s)} =  \cD_{(\a_1}{}^{\bd_1} \dots \cD_{\a_s}{}^{\bd_s} h_{\a_{s+1} \dots \a_{2s}) \bd(s)} ~.
\ee
Further, $\mathcal{C}_{\a(2s)}$ satisfies the Bianchi identity \eqref{2.4}
\be
\cD^{\b_1}{}_{(\ad_1} \dots \cD^{\b_s}{}_{\ad_s)} \mathcal{C}_{\a(s) \b(s)} 
=
\cD_{(\a_1}{}^{\bd_1} \dots \cD_{\a_s)}{}^{\bd_s} \bar{\mathcal{C}}_{\ad(s) \bd(s)} ~.
\ee

Next, the geometry of the $\N=1$ AdS superspace, $\text{AdS}^{4|4}$, is described by the covariant derivatives $\cD_{A} = (\cD_a, \cD_\a,\bar{\cD}^\ad)$, which obey the algebra (see, e.g., \cite{BK})
\begin{subequations}
\bea
\{ \cD_\a , \cD_\b \} &=& - 4 \bar{\mu} M_{\a \b} ~, \quad 
 \{ \cDB_\ad , \cDB_\bd \} = 4 \mu \bar{M}_{\ad\bd} ~, \quad
\{ \cD_\a , \bar{\cD}_{\ad} \} = 2 \ri \cD_\aa ~, \\
\left[ \cD_\a , \cD_{\b \bd} \right] &=& \ri \mub \ve_{\a\b} \cDB_\bd \ , \quad [\cDB_\ad , \cDB_{\b\bd}] 
= - \ri \mu \ve_{\ad\bd} \cD_\b ~,
\eea
\end{subequations}
where $\mu$ is a nonzero constant having the dimension of mass. 
In this supergeometry the field strengths $\cW_{\a(2s+1)}$ take the form \cite{KS94}
\be
\mathcal{W}_{\a(2s+1)} = - \frac{1}{4} (\bar{\cD}^2 - 4 \mu) \cD_{(\a_1}{}^{\bd_1} \dots \cD_{\a_s}{}^{\bd_s} \cD_{\a_{s+1}} H_{\a_{s+2} \dots \a_{2s+1}) \bd(s)} ~,
\ee
and obey the Bianchi identity
\be
\ri \cD^{\b_1}{}_{(\ad_1} \dots \cD^{\b_s}{}_{\ad_s)} \cD^{\b_{s+1}} \mathcal{W}_{\a(s) \b(s+1)} 
=
- \ri \cD_{(\a_1}{}^{\bd_1} \dots \cD_{\a_s)}{}^{\bd_s} \bar{\cD}^{\bd_{s+1}} \bar{\mathcal{W}}_{\ad(s) \bd(s+1)}  ~.
\ee

Finally, we consider $\text{AdS}^{4|8}$, the $\N=2$ AdS superspace. Recall that its geometry is encoded within the covariant derivatives $\cD_A = (\cD_a, \cD_\a^i, \bar \cD^\ad_i)$, which obey the algebra (see \cite{KT-M-ads} for the technical details):
\be
\{ \cD_\a^i , \cD_\b^j \} = 4 S^{ij} M_{\a \b} + 2 \ve_{\a \b} \ve^{i j} S^{kl} J_{kl} ~, \quad \{ \cD_\a^i , \bar \cD^\bd_j \} = - 2 \ri \d_j^i \cD_{\a}{}^{\bd} ~,
\ee
where ${S}^{ij} $ is a nonzero covariantly constant
iso-triplet, ${S}^{ji} = { S}^{ij}$, satisfying  the integrability condition $[S, S^\dagger ]=0$, 
with $S= (S^i{}_j)$.\footnote{The integrability condition implies that $S^{ij}$ can be chosen to be real, $\overline{ {S}^{ij}} = {S}_{ij} =\ve_{ik}\ve_{jl}{ S}^{kl}$. However we will not impose the reality condition.} Here, the field strengths $\bm{\cW}_{\a(2s+2)}$ take the form
\be
\bm{\cW}_{\a(2s+2)} = \frac{1}{48} (\bar \cD^{ij} + 4 \bar S^{ij})\bar \cD_{ij} \cD_{(\a_1}{}^{\bd_1} \dots \cD_{\a_s}{}^{\bd_s} \cD_{\a_{s+1} \a_{s+2}} H_{\a_{s+3} \dots \a_{2s+2}) \bd(s)} ~,
\ee
and satisfy the Bianchi identity
\be
\cD^{\b_1}{}_{(\ad_1} \dots \cD^{\b_s}{}_{\ad_s)} \cD^{\b_{s+1} \b_{s+2}} \bm{\mathcal{W}}_{\a(s) \b(s+2)} 
=
\cD_{(\a_1}{}^{\bd_1} \dots \cD_{\a_s)}{}^{\bd_s} \bar{\cD}^{\bd_{s+1} \bd_{s+2}} \bar{\bm{\mathcal{W}}}_{\ad(s) \bd(s+2)}  ~.
\ee

Throughout this work we have restricted our attention to the analysis of models described by real gauge prepotentials. However, our construction readily extends itself to the complex case. In particular, one may consider a complex CHS gauge prepotential $\phi_{\a(m)\ad(n)}$, with $m,n \geq 1$ and $m \neq n$ 
\cite{KP,KMT,Vasiliev2009}. It is a primary field, $K_b \phi_{\a(m)\ad(n)} = 0$, defined modulo
\be
\label{ComplexGT}
\d_\ell \phi_{\a(m)\ad(n)} = \nabla_{(\a_1 (\ad_1} \ell_{\a_2 \dots \a_m) \ad_2 \dots \ad_n)} ~,
\ee
where the gauge parameter $\ell_{\a(m-1)\ad(n-1)}$ is also primary. Conformal invariance of \eqref{ComplexGT} uniquely fixes the dimension of the gauge field, $\mathbb{D} \phi_{\a(m)\ad(n)} = (2 - \hf(m+n)) \phi_{\a(m)\ad(n)}$. The action functional describing its dynamics is required to take the form $S^{(m,n)}[\hat{\cC}, \check{\cC},\bar{\hat{\cC}},\bar{\check{\cC}}]$, where we have made the definitions
\begin{subequations}
	\label{ComplexFS}
\begin{align}
	\hat{\cC}_{\a(m+n)} &= \nabla_{(\a_1}{}^{\bd_1} \dots \nabla_{\a_n}{}^{\bd_n} \phi_{\a_{n+1} \dots \a_{m+n}) \bd(n)} ~, \\
	\check{\cC}_{\a(m+n)} &= \nabla_{(\a_1}{}^{\bd_1} \dots \nabla_{\a_m}{}^{\bd_m} \bar{\phi}_{\a_{m+1} \dots \a_{m+n}) \bd(m)} ~.
\end{align}
\end{subequations}
The descendants introduced above are primary in generic backgrounds, 
\begin{subequations} 
\bea 
K_b \hat{\cC}_{\a(m+n)} &=&0~, \qquad 
\mathbb{D} \hat{\cC}_{\a(m+n)} = \Big(2 + \hf(n-m)\Big) \hat{\cC}_{\a(m+n)}~;\\
K_{b}\check{{\cC}}_{\a(m+n)}&=&0~,\qquad \mathbb{D}\check{{\cC}}_{\a(m+n)}
=\Big(2+\frac{1}{2}(m-n)\Big)\check{{\cC}}_{\a(m+n)}~.
\eea
\end{subequations} 
 and gauge-invariant in all conformally flat ones, 
 \bea
 C_{\a(4)} =0 \quad \implies \quad \d_\ell \hat{\cC}_{\a(m+n)} = \d_\ell  \check{\cC}_{\a(m+n)} =0~.
 \eea

It is important to note that the field strengths \eqref{ComplexFS} also obey the Bianchi identity
\be
\label{ComplexBI}
\nabla^{\b_1}{}_{(\ad_1} \dots \nabla^{\b_m}{}_{\ad_m)} \hat{\cC}_{\a(n) \b(m)} 
= \nabla_{(\a_1}{}^{\bd_1} \dots \nabla_{\a_n)}{}^{\bd_n} \bar{\check{\cC}}_{\ad(m) \bd(n)} ~.
\ee
Now, considering $S^{(m,n)}[\hat{\cC},\check{\cC},\bar{\hat{\cC}},\bar{\check{\cC}}]$ as a functional of the unconstrained fields $\hat{\cC}_{\a(m+n)}$, $\check{\cC}_{\a(m+n)}$ and their conjugates, we may introduce the primary fields
\be
\ri^{m+n+1} \hat{\cM}_{\a(m+n)} :=  \frac{\d S^{(m,n)}[\hat{\cC},\check{\cC},\bar{\hat{\cC}},\bar{\check{\cC}}]}{\d \check{\cC}^{\a(m+n)}} ~, 
\quad \ri^{m+n+1} \check{\cM}_{\a(m+n)} :=  \frac{\d S^{(m,n)}[\hat{\cC},\check{\cC}~,\bar{\hat{\cC}},\bar{\check{\cC}}]}{\d \hat{\cC}^{\a(m+n)}}~,
\label{5.15}
\ee
where we have made the definition
\begin{align}
\d S^{(m,n)}[\hat{\cC},\check{\cC},\bar{\hat{\cC}},\bar{\check{\cC}}] &= \int \rd^4x\, e \, 
\Big \{ \d \hat{\mathcal{C}}^{\a(m+n)} \frac{\d S^{(m,n)}[\hat{\cC},\check{\cC},\bar{\hat{\cC}},\bar{\check{\cC}}]}{\d \hat{\mathcal{C}}^{\a(m+n)}} \non \\
& \qquad \qquad \qquad \qquad + \d \check{\mathcal{C}}^{\a(m+n)} \frac{\d S^{(m,n)}[\hat{\cC},\check{\cC},\bar{\hat{\cC}},\bar{\check{\cC}}]}{\d \check{\mathcal{C}}^{\a(m+n)}} \Big \} + \text{c.c.}
\end{align}
The conformal properties of the fields \eqref{5.15} are: 
\begin{subequations} 
\bea 
K_b \hat{\cM}_{\a(m+n)} &=&0~, \qquad 
\mathbb{D} \hat{\cM}_{\a(m+n)} = \Big(2 + \hf(n-m)\Big) \hat{\cM}_{\a(m+n)}~;\\
K_{b}\check{{\cM}}_{\a(m+n)}&=&0~,\qquad \mathbb{D}\check{{\cM}}_{\a(m+n)}
=\Big(2+\frac{1}{2}(m-n)\Big)\check{{\cM}}_{\a(m+n)}~.
\eea
\end{subequations} 
Varying $S^{(m,n)}[\hat{\cC},\check{\cC},\bar{\hat{\cC}},\bar{\check{\cC}}]$ with respect to $\phi_{\a(m) \ad(n)}$ yields
\bea
\label{ComplexEoM}
\nabla^{\b_1}{}_{(\ad_1} \dots \nabla^{\b_m}{}_{\ad_m)} \hat{\cM}_{\a(n) \b(m)} 
= \nabla_{(\a_1}{}^{\bd_1} \dots \nabla_{\a_n)}{}^{\bd_n} \bar{\check{\cM}}_{\ad(m) \bd(n)} ~.
\eea

It is clear from the discussion above that the system of equations \eqref{ComplexBI} and \eqref{ComplexEoM} are invariant under the $\sU(1)$ duality rotations
\begin{subequations}
\begin{align}
	\d_\l \hat{\cC}_{\a(m+n)} = \l \hat{\cM}_{\a(m+n)} ~, \quad \d_\l \check{\cC}_{\a(m+n)} = \l \check{\cM}_{\a(m+n)} ~, \\
	\d_\l \hat{\cM}_{\a(m+n)} = - \l \hat{\cC}_{\a(m+n)} ~, \quad \d_\l \check{\cM}_{\a(m+n)} = - \l \check{\cC}_{\a(m+n)} ~.
\end{align}
\end{subequations}
One may then perform similar analyses to those undertaken in section \ref{section2} and construct $\sU(1)$ duality-invariant nonlinear models for such fields.\footnote{We mention in passing that such a construction can also be uplifted to the case of a SCHS theory described by a complex prepotential, such as those of \cite{KP,KR19,KPR}.} 
They satisfy  the self-duality equation
\bea
\ri^{m+n+1} \int \rd^4x \, e \, \Big \{ \hat{\cC}^{\a(m+n)}  \check{\cC}_{\a(m+n)}
+ \hat{\cM}^{\a(m+n)} \check{\cM}_{\a(m+n)} \Big \} + \text{c.c.}  = 0 ~,
\eea
which must hold for unconstrained fields $\hat{\cC}_{\a(m+n)}$ and $\check{\cC}_{\a(m+n)}$.
The simplest solution of this equation is the free CHS action 
\bea
S^{(m,n)} [\hat{\cC},\check{\cC},\bar{\hat{\cC}},\bar{\check{\cC}}]
= {\ri^{m+n}}\int \rd^4 x \, e\,  \hat{ {\cC}}^{\a(m+n)}\check{{\cC}}_{\a(m+n)} 
+{\rm c.c.}~, 
\eea
which reduces to \eqref{D.1} for  $m=n=s$ and real prepotential.

$\sU(1)$ duality-invariant actions of the type considered above naturally arise at the component level of supersymmetric self-dual theories, such as those discussed in sections \ref{section3} and \ref{section4}. Such considerations are beyond the scope of the current paper and present an interesting avenue for further work. It would also be interesting to study duality-invariant models of the higher-depth (super)fields of \cite{KP,KPR2}.
\\


\noindent
{\bf Acknowledgements:}\\
We thank the referee for useful comments and suggestions.
SK is grateful to Stefan Theisen and Arkady Tseytlin for discussions. 
His work is supported in part by the Australian 
Research Council, project No. DP200101944.
The work of ER is supported by the Hackett Postgraduate Scholarship UWA,
under the Australian Government Research Training Program.

 
\appendix

\section{Conformal space}\label{appendixA}

This appendix reviews the salient details of conformal geometry in four dimensions 
\cite{KTvN1} 
pertinent to this work.\footnote{See also \cite{FT} for a pedagogical review of conformal (super)gravity and \cite{KP,ButterN=1,BKNT-M1} for the modern formulation of conformal geometry we use.}
 We adopt the spinor conventions of \cite{BK}, which are similar to those of \cite{WB}. We consider a curved spacetime $\mathcal{M}^{4}$ parametrised by local coordinates $x^m$. The structure group is chosen to be $\sSU(2,2)$, whose Lie algebra is spanned by the translation $P_a$, Lorentz $M_{ab}$, dilatation $\mathbb{D}$ and the special conformal $K_a$ generators. The covariant derivatives $\nabla_a$ then take the form
\begin{align}
\label{A.1}
\nabla_{a}=e_{a}{}^{m}\partial_m-\frac{1}{2}\omega_{a}{}^{bc}M_{bc}-\mathfrak{b}_a\mathbb{D}-\mathfrak{f}_{a}{}^{b}K_b~,
\end{align}
where $e_{a}{}^{m}$ is the inverse vielbein, $\omega_{a}{}^{bc}$ the Lorentz spin connection, $\mathfrak{b}_a$ the dilatation connection and $\mathfrak{f}_{a}{}^{b}$ the special conformal connection. The commutation relations of $\nabla_a$ with 
the generators  $M_{ab}$, $\mathbb{D}$ and $K_a$ are obtained from those of 
$P_a$ with the same generators by replacing $P_a \to \nabla_a$. 

The covariant derivatives \eqref{A.1} obey the commutation relations
\bea
\label{A.2}
\big[\nabla_{\a\ad},\nabla_{\b\bd} \big]&=&-\big(\ve_{\ad\bd}C_{\a\b\g\d}M^{\g\d}+\ve_{\a\b}\bar{C}_{\ad\bd\gd\dd}\bar{M}^{\gd\dd}\big) \notag\\
&&
-\frac{1}{4}\big(\ve_{\ad\bd}\nabla^{\d\gd}C_{\a\b\d}{}^{\g}+\ve_{\a\b}\nabla^{\g\dd}\bar{C}_{\ad\bd\dd}{}^{\gd}\big)K_{\g\gd}~.
\eea
Here $C_{\a\b\g\d}$ and $\bar{C}_{\ad\bd\gd\dd}$ are the self-dual and anti self-dual parts of the Weyl tensor $C_{abcd}$, and are primary. We remind the reader that a field $\varphi$ (with suppressed indices) is said to be primary if it obeys 
\be
K_\aa \varphi = 0~.
\ee
The commutation relations \eqref{A.2} should be accompanied by the relations
\bea
\big[\mathbb{D},\nabla_{\aa} \big]=\nabla_{\aa}~,\qquad \big[K_{\a\ad},\nabla_{\b\bd}\big] &=& 4\big(\ve_{\ad\bd}M_{\a\b}+\ve_{\a\b}\bar{M}_{\ad\bd}-\ve_{\a\b}\ve_{\ad\bd}\mathbb{D}\big)~,
\eea
and we recall that the Lorentz generators act on vectors and Weyl spinors as follows:
\bea
M_{ab} V_{c} = 2 \eta_{c[a} V_{b]} ~, \qquad 
M_{\a \b} \j_{\g} = \ve_{\g (\a} \j_{\b)} ~, \qquad \bar{M}_{\ad \bd} \bar \j_{\gd} = \ve_{\gd ( \ad} \bar \j_{\bd )} ~.
\eea


\section{$\N=1$ conformal superspace}\label{appendixB}

In this appendix we review the elements of the  $\N=1$ conformal superspace approach to off-shell conformal supergravity relevant to this work. For more details
we refer the reader to the original paper \cite{ButterN=1} (see also appendix A of \cite{KPR}).

Consider a curved $\cN=1$ superspace $\mathcal{M}^{4|4}$
parametrised by local coordinates 
$z^{M} = 
(x^{m},\theta^{\m},\bar \theta_{\dot{\mu}})$.  
The structure group is chosen to be $\sSU(2,2|1)$. Its corresponding superalgebra is spanned by the translation $P_A=(P_a, Q_\a ,\bar Q^\ad)$, Lorentz $M_{ab}$, dilatation $\mathbb{D}$,  R-symmetry $Y$ and the special conformal $K^A=(K^a, S^\a ,\bar S_\ad)$ generators. The covariant derivatives 
$\nabla_A$
then have the form
\begin{align}
	\nabla_A &= (\nabla_a, \nabla_\alpha, \bar\nabla^\ad)	=E_A{}^M \pa_M - \hf \Omega_A{}^{bc} M_{bc} - \ri \Phi_A Y
	- B_A \mathbb{D} - \mathfrak{F}_{AB}K^B ~,
	\label{B.1}
\end{align}
where $E_{A}{}^{M}$ denotes the superspace inverse vielbein, $\Omega_A{}^{bc}$ the Lorentz connection,  $\Phi_A$  the  $\rm U(1)_R$ connection, $B_A$
the dilatation connection, and  $\mathfrak F_{AB}$ the special
superconformal connection.

The covariant derivatives \eqref{B.1} obey the algebra
\begin{subequations}
	\label{CSSAlgebra}
	\bea
	\{ \nabla_{\a} , \nabla_{\b} \} & = & 0 ~, \quad \{\nabla_{\a} , \bar{\nabla}_{\ad} \} = - 2 \ri \nabla_{\a \ad} ~, \\
	\big[ \bar{\nabla}_{\ad} , \nabla_{\b \bd} \big] & = & - \ri \ve_{\ad \bd} \Big( 2 W_{\b}{}^{\g \d} M_{\g \d} + \frac{1}{2} \nabla^{\a} W_{\a \b \g} S^{\g} + \frac{1}{2} \nabla^{\a \gd} W_{\a \b}{}^{\g} K_{\g \gd} \Big) ~,
	\eea
\end{subequations}
Here $W_{\a \b \g}$ is the $\N=1$ super-Weyl tensor and is subject to the constraints: 
\begin{subequations}
	\be
	K^D W_{\a \b \g} =0~, \quad \bar \nabla^\dd W_{\a\b\g}=0 ~, \quad 
	{\mathbb D} W_{\a\b\g} = \frac 32 W_{\a\b\g}~, \quad YW_{\a\b\g}=-W_{\a\b\g}~, 
	\ee
	as well as the Bianchi identity
	\be
	\label{B.3b}
	B_{\a\ad} :=  \ri \nabla^\b{}_{\ad} \nabla^\g W_{\a\b\g}
	=\ri \nabla_{\a}{}^{ \bd} \bar \nabla^\gd \bar W_{\ad\bd\gd}
	= \bar B_{\a\ad}~,
	\ee
\end{subequations}
where the primary superfield $B_{\a\ad}$ is the super-Bach tensor and was introduced in \cite{BK88}, see also \cite{KMT,KP,BK}.

We remind the reader that a tensor superfield $\J$ (with suppressed indices) is said to be primary 
and of dimension $\D_\J$ and $\sU(1)_R$ charge $q_\J$ if the following conditions hold:
\bea
K^B \Psi = 0 ~, \qquad  \mathbb{D} {\Psi} = \D_{{\Psi}} {\Psi} ~, \qquad 
Y {\Psi} = q_{{\Psi}} {\Psi} ~.
\eea
Of particular importance are primary chiral superfields, which satisfy
\bea
K^B \Psi = 0 ~, \qquad \bar{\nabla}^{\bd} {\J} = 0~.
\eea
Requiring consistency of these constraints with the superconformal algebra yields highly non-trivial implications. Specifically, it must take the form $\Psi = \Psi_{\a(m)}$, and its $\sU(1)_R$ charge and dimension are related as follows:
\bea
\label{chiralDimChargeN=1}
q_\Psi = - \frac{2}{3} \D_\Psi ~.
\eea

The algebra \eqref{CSSAlgebra} is to be accompanied by the following (anti-)commutation relations: the  $\rm U(1)_R$, dilatation and special conformal generators obey
\begin{subequations}
	\begin{align}
		[Y, \nabla_\a] &= \nabla_\a ~,\quad [Y, \bar\nabla^\ad] = - \bar\nabla^\ad~,   \\
		[\mathbb{D}, \nabla_\aa] &= \nabla_\aa ~, \quad
		[\mathbb{D}, \nabla_\a] = \hf \nabla_\a ~, \quad
		[\mathbb{D}, \bar\nabla^\ad ] = \hf \bar\nabla^\ad ~\\
		[Y, S^\a] &= - S^\a ~, \quad
		[Y, \bar{S}_\ad] = \bar{S}_\ad~, \quad \{ S_\a , \bar{S}_\ad \} = 2 \ri  K_{\aa} \\
		[\mathbb{D}, K_\aa] &= - K_\aa ~, \quad
		[\mathbb{D}, S^\a] = - \hf S^\a~, \quad
		[\mathbb{D}, \bar{S}_\ad ] = - \hf \bar{S}_\ad ~,
	\end{align}
	while the algebra of $K_A$ and $\nabla_B$ takes the form
	\begin{align}
		[K_\aa, \nabla_\bb] &= 4 \big(\ve_{\ad \bd} M_{\a \b} +  \ve_{\a \b} \bar{M}_{\ad \bd} -  \ve_{\a \b} \ve_{\ad \bd} \mathbb{D} \big) ~, \\
		\{ S_\a , \nabla_\b \} &= \ve_{\a \b} \big( 2 \mathbb{D} - 3 Y \big) - 4 M_{\a \b} ~, \\
		\{ \bar{S}_\ad , \bar{\nabla}_\bd \} &= - \ve_{\ad \bd} \big( 2 \mathbb{D} + 3 Y) + 4 \bar{M}_{\ad \bd}  ~, \\
		[K_{\a \ad}, \nabla_\b] &= - 2 \ri \ve_{\a \b} \bar{S}_{\ad} \ , \qquad \qquad \qquad[K_\aa, \bar{\nabla}_\bd] =
		2 \ri  \ve_{\ad \bd} S_{\a} ~,  \\
		[S_\a , \nabla_\bb] &= 2 \ri \ve_{\a \b} \bar{\nabla}_{\bd} \ , \qquad \qquad \quad \qquad[\bar{S}_\ad , \nabla_\bb] =
		- 2 \ri \ve_{\ad \bd} \nabla_{\b} ~,
	\end{align}
\end{subequations}
where all other graded commutators vanish.

The chiral and full superspace integrals are related according to the rule
\bea
 \int \rd^4x\rd^2\theta\rd^2\bar\theta\, E\,  U
 =- \frac 14 \int \rd^4x \rd^2\theta\, \cE\, \bar \nabla^2 U~,\qquad E^{-1} = {\rm Ber}(E_A{}^M)~,
 \label{B.8}
\eea
where $U$ is a primary real  superfield of dimension $+2$.


\section{$\mathcal{N}=2$ conformal superspace}\label{appendixC}

This appendix reviews $\mathcal{N} = 2$ conformal superspace, a formulation for off-shell $\mathcal{N}=2$ conformal supergravity developed by Butter \cite{ButterN=2} and then reformulated in \cite{BN}. 

We consider a curved $\cN=2$ superspace $\mathcal{M}^{4|8}$ parametrised by local coordinates 
$z^{M} = (x^{m},\theta^{\m}_\imath,\bar \theta_{\dot{\mu}}^\imath)$.  
The structure group is chosen to be $\sSU(2,2|2)$. The corresponding superalgebra is spanned by the Lorentz $M_{ab}$, translation $P_A=(P_a, Q_\a^i ,\bar Q^\ad_i)$, dilatation $\mathbb{D}$,  R-symmetry $Y$ and $J_{ij}$, and the special conformal $K^A=(K^a, S^\a_i ,\bar S_\ad^i)$ generators. The covariant derivatives $\nabla_A = (\nabla_a, \nabla_\alpha^i, \bar\nabla^\dalpha_i)$ then have the form
\begin{align}
	\nabla_A &= E_A - \hf \Omega_A{}^{ab} M_{ab} - \Phi_A{}^{ij} J_{ij} - \ri \Phi_A Y
	- B_A \mathbb{D} - \frak{F}_{AB} K^B ~.
	\label{C.1}
\end{align}
As compared with \eqref{B.1}, we have introduced $\Phi_A{}^{ij}$, the $\sSU(2)_R$ connection. The corresponding generator acts on isospinors as follows:
\be
J^{ij} \chi^k = \ve^{k(i} \chi^{j)} ~.
\ee

The covariant derivatives \eqref{C.1} obey the algebra
\begin{subequations}\label{CSGAlgebra}
	\begin{align}
		\{ \nabla_\a^i , \nabla_\b^j \} &= 2 \ve^{ij} \ve_{\a\b} \bar{W}_{\gd\dd} \bar{M}^{\gd\dd} + \hf \ve^{ij} \ve_{\a\b} \bar{\nabla}_{\gd k} \bar{W}^{\gd\dd} \bar{S}^k_\dd - \hf \ve^{ij} \ve_{\a\b} \nabla_{\g\dd} \bar{W}^\dd{}_\gd K^{\g \gd}~, \\
		\{ \nabla_\a^i , \bar{\nabla}^\bd_j \} &= - 2 \ri \d_j^i \nabla_\a{}^\bd~, \\
		[\nabla_{\a\ad} , \nabla_\b^i ] &= - \ri \ve_{\a\b} \bar{W}_{\ad\bd} \bar{\nabla}^{\bd i} - \frac{\ri}{2} \ve_{\a\b} \bar{\nabla}^{\bd i} \bar{W}_{\ad\bd} \mathbb{D} - \frac{\ri}{4} \ve_{\a\b} \bar{\nabla}^{\bd i} \bar{W}_{\ad\bd} Y + \ri \ve_{\a\b} \bar{\nabla}^\bd_j \bar{W}_{\ad\bd} J^{ij}
		\eol & \quad
		- \ri \ve_{\a\b} \bar{\nabla}_\bd^i \bar{W}_{\gd\ad} \bar{M}^{\bd \gd} - \frac{\ri}{4} \ve_{\a\b} \bar{\nabla}_\ad^i \bar{\nabla}^\bd_k \bar{W}_{\bd\gd} \bar{S}^{\gd k} + \frac{1}{2} \ve_{\a\b} \nabla^{\g \bd} \bar{W}_{\ad\bd} S^i_\g
		\eol & \quad
		+ \frac{\ri}{4} \ve_{\a\b} \bar{\nabla}_\ad^i \nabla^\g{}_\gd \bar{W}^{\gd \bd} K_{\g \bd}~.
	\end{align}
\end{subequations}	
\begin{subequations}
	Here $W_{\a \b}$ is the $\N=2$ super-Weyl tensor and is subject to the constraints:
	\bea
	K^C W_{\a \b}  = 0 ~, \quad \bar{\nabla}^\gd_k W_{\a \b} = 0 ~, \quad \mathbb{D} W_{\a \b} = W_{\a \b}, \quad Y W_{\a \b} = - W_{\a \b} ~.
	\eea
	We also find that $W_{\a \b}$ obeys the Bianchi identity
	\begin{align}
		B = \nabla_{\a\b} W^{\a\b} &= \bar{\nabla}^{\ad\bd} \bar{W}_{\ad\bd}  = \bar{B} ~,
	\end{align}
	\end{subequations}
	where the primary superfield $B$ is the $\N=2$ super-Bach tensor. 
	We remind the reader that a superfield $\Phi$ (with suppressed indices) is said to be primary of dimension $\D_\J $ and $\sU(1)_R$ charge $q_\J$ if the following conditions hold:
	\bea
K^B \Psi = 0 ~, \qquad \mathbb{D} \Psi = \D_\Psi \Psi ~, \qquad Y \Psi = q_\Psi \Psi ~.
\eea
Of particular importance are primary chiral superfields, which satisfy
\bea
K^B \Psi = 0 ~, \qquad \bar{\nabla}^{\bd}_j \J = 0~.
\eea
The consistency of these constraints with the superconformal algebra leads to highly non-trivial implications. In particular, it can carry no isospinor or dotted spinor indices, $\Psi = \Psi_{\a(m)}$, and its $\sU(1)_R$ charge and dimension are related as follows:
\bea
\label{chiralDimCharge}
q_\Psi = - 2 \D_\Psi ~.
\eea
Further, we note that for any primary tensor superfield ${\mathfrak U}_{\a(m)}$ with the property $q_{\mathfrak U} = - 2 \D_{\mathfrak U}$, the following object 
\bea
\Psi_{\a(m)} = \bar{\nabla}^4 {\mathfrak U}_{\a(m)} \equiv \frac{1}{48} \bar{\nabla}^{ij} \bar{\nabla}_{ij} {\mathfrak U}_{\a(m)}
\eea
is both primary and chiral in conformally flat backgrounds \cite{ButterN=2,KTM08}. 

The chiral and full superspace integrals are related according to the rule
\bea
 \int \rd^4x\rd^4\theta\rd^4\bar\theta\, \bm{E}\,  U
 =\int \rd^4x \rd^4\theta\, \bm{\cE}\, \bar \nabla^4 U~,\qquad \bm{E}^{-1} = {\rm Ber}(E_A{}^M)~,
 \label{C.9}
\eea
where $U$ is a primary real dimension-0 superfield.

For further details regarding $\cN=2$ conformal superspace, we refer the reader to the original work \cite{ButterN=2}, as well as \cite{BN,KR21}.
%


\section{Elimination of auxiliary variables} \label{appendixNew}

This appendix is devoted to a derivation of the conformal $\sU(1)$ duality-invariant CHS models \eqref{2.13} via the auxiliary variable formalism introduced in section \ref{section2.3}. We emphasise that the latter is a higher-spin generalisation of the Ivanov-Zupnik approach \cite{IZ_N3,IZ1,IZ2}. 

Consider the following action functional
\begin{align}
	\label{AuxAction}
	S^{(s)}[\mathcal{C},\bar{\mathcal C}, \r, \bar \r] = (-1)^s \int \rd^4x 
	\, e \, \Big \{ 2 \r \mathcal{C} - 
	\r^2- \frac{1}{2}\mathcal{C}^2 \Big \} + \text{c.c.} + \b \int \rd^4x 
	\, e \, \sqrt{\r^2 \bar{\r}^2} ~,
\end{align} 
where one should keep in mind the definitions of section \ref{section2}. It is clear that \eqref{AuxAction} is both conformal and $\sU(1)$ duality-invariant. Varying this action with respect to the auxiliary variable $\r^{\a(2s)}$ yields
\begin{align}
	\r_{\a(2s)} = \cC_{\a(2s)} + \frac{(-1)^s \b}{2} \frac{\r_{\a(2s)} \bar{\r}^2}{\sqrt{\r^2 \bar{\r}^2}} ~.
\end{align}
Employing this result, it is possible to integrate out $\r_{\a(2s)}$. As a result, we obtain the self-dual model
\begin{align}
	\label{SDAction}
	S^{(s)}[\mathcal{C},\bar{\mathcal C}] = \frac{(-1)^s}{2} \frac{1 + (\b/2)^2}{1-(\b/2)^2} \int \rd^4x \, e \, \Big \{ \mathcal{C}^2 + \bar{\mathcal{C}}^2 \Big \} +  \, \frac{\b}{1 - (\b/2)^2} \int \rd^4x \, e \, \sqrt{\mathcal{C}^2 \bar{\mathcal{C}}^2} ~.
\end{align}
Now, upon making the identification
\bea
\cosh \g = \frac{1 + (\b/2)^2}{1-(\b/2)^2}  \quad \Longleftrightarrow \quad \sinh \g = \frac{\b}{1-(\b/2)^2} ~,
\eea
it is clear that \eqref{SDAction} coincides with \eqref{2.13}, which concludes our analysis. It is important to note that, in the $s=1$ case, this computation was first performed in \cite{K21}.


\section{Overall signs for free (S)CHS actions} \label{appendixD}

In this appendix we show that the overall signs of the free CHS actions 
\cite{FT,FL,FL2} 
\bea \label{D.1}
S^{(s)}[\mathcal{C},\bar{\mathcal C}] = \frac{(-1)^s  }{2}  \int \rd^4x \, e \, 
\Big \{ \mathcal{C}^{\a(2s)} \mathcal{C}_{\a(2s)} + {\rm c.c.}  \Big \}
\eea
can be fixed by making use of supersymmetry considerations in conjunction
with the known action of Maxwell theory 
for $s=1$,
\bea 
S^{(1)}[\mathcal{C},\bar{\mathcal C}] = -\hf  \int \rd^4x \, e \, 
\Big \{ \mathcal{C}^{\a(2)} \mathcal{C}_{\a(2)} + {\rm c.c.}  \Big \}
= -\frac 14 \int \rd^4x \, e \, C^{ab}C_{ab}~,
\eea
where $C_{ab} =\nabla_a h_b - \nabla_b h_a$. Moreover, similar arguments
allow us to correctly fix the overall signs of the free $\cN=1$ and $\cN=2$ SCHS actions. 

The overall sign in \eqref{D.1} is also fixed by identifying the action $S^{(s)}[\mathcal{C},\bar{\mathcal C}] $ with the induced one obtained by computing the logarithmically divergent part of the effective action of a conformal scalar field coupled 
to background conformal higher-spin fields \cite{BT}.

\subsection{$\N=1$ actions}

Consider the chiral field strength $\cW_{\a(2s+1)}$ and introduce its bosonic components 
\begin{subequations}
\label{D.2}
\begin{align}
\cC_{\a(2s)} &:= \frac{1}{4} \sqrt{\frac{2s+1}{s+1}} \nabla^\b \cW_{\b \a(2s) } |_{\theta = 0} ~, \\
\cC_{\a( 2s+2)} &:= \frac{\sqrt 2 \ri }{4} \nabla_{(\a_1 } \cW_{\a_2 \dots  \a_{2s+2}) } |_{\theta = 0}~, 
\end{align}
\end{subequations}
which have been defined such that Bianchi identity \eqref{2.4} holds both for field strengths. We then compute the bosonic part of the SCHS action \cite{KMT,KP}
\bea \label{D.3}
S^{(s)} [\mathcal{W},\bar{\mathcal W}] &:=& \frac{z_s}{4} \int \rd^4x 
	\rd^2 \q \, \mathcal{E} \, \mathcal{W}^{\a(2s+1)} \cW_{\a(2s+1)}+ \text{c.c.} \non \\
	&=&- \frac{z_s}{16}  \int \rd^4x \, e
	 \, \nabla^2\Big( \mathcal{W}^{\a(2s+1)} \cW_{\a(2s+1)}\Big) \Big |_{\theta = \bar{\theta} = 0} + \text{c.c.} \non \\
	&=& z_s \int \rd^4x \, e \, \Big (
	 \cC^{\a(2s)} \mathcal{C}_{\a(2s)} - \cC^{\a(2s+2)} \cC_{\a(2s+2)} \Big ) + \dots + \text{c.c.}
\eea 
where the ellipses denote the fermionic sector, which is irrelevant to our analysis. For the overall signs of the component actions present in \eqref{D.3} to agree with those of \eqref{D.1}, we require $z_s = (-1)^{s}$. Therefore, the $\N=1$ SCHS actions take the form
\cite{KP,KMT}
\bea
\label{D.5}
S^{(s)} [\mathcal{W},\bar{\mathcal W}] = \frac{(-1)^s}{4} \int \rd^4x 
\rd^2 \q \, \mathcal{E} \, \mathcal{W}^{\a(2s+1)} \cW_{\a(2s+1)}+ \text{c.c.}
\eea


\subsection{$\N=2$ actions}

Consider the chiral field strength $\bm{\mathcal{W}}_{\a(2s+2)}$. It contains two $\N=1$ fermionic superfields in its multiplet
\begin{subequations}
\label{D.6}
\begin{align}
\cW_{\a(2s+1)} &:= \frac{\ri}{2} \sqrt{\frac{s+1}{2s+3}} \nabla^{\b \underline{2}} \bm{\cW}_{\a(2s+1) \b} |_{\theta_{\underline{2}} = 0} ~,  \\
\cW_{\a(2s+3)} &:= \frac{\sqrt 2}{4} \nabla_{(\a_1}^{\underline 2} \bm{\cW}_{\a_2 \dots \a_{2s+3})}|_{\theta_{\underline{2}} = 0}~,
\end{align}
\end{subequations}
which have been defined such that the Bianchi identity \eqref{3.4} holds for both field strengths. Next, we reduce the $\cN=2$ SCHS action \cite{KR21} to $\N=1$ superspace
\bea \label{D.7}
S^{(s)} [\bm{\mathcal{W}},\bar{\bm{\mathcal W}}] &:=& \frac{z_s}{4} \int \rd^4x 
\rd^4 \q \, \bm{\mathcal{E}} \, \bm{\mathcal{W}}^{\a(2s+1)} \bm{\cW}_{\a(2s+1)}+ \text{c.c.} \non \\
&=&- \frac{z_s}{16}  \int \rd^4x 
\rd^2 \q \, \mathcal{E} \, \nabla^{\a \underline{2}} \nabla_{\a}^{\underline 2} \Big( \bm{\mathcal{W}}^{\a(2s+1)} \bm{\cW}_{\a(2s+1)}\Big) \Big |_{\theta^{\underline{2}} = \bar{\theta}_{\underline{2}} = 0} + \text{c.c.} \non \\
&=& z_s \int \rd^4x 
\rd^2 \q \, \mathcal{E} \, \Big (
\cW^{\a(2s+1)} \cW_{\a(2s+1)} - \cW^{\a(2s+3)} \cW_{\a(2s+3)} \Big ) \non \\
&&+ \dots + \text{c.c.} 
\eea 
where the ellipses denotes the sector containing integer superspin field strengths, which is irrelevant to our analysis. For the overall signs of the $\N=1$ actions present in \eqref{D.7} to agree with those of \eqref{D.5}, we require $z_s = (-1)^{s}$. Therefore, the $\N=2$ SCHS actions take the form
\cite{KR21}
\bea
S^{(s)} [\bm{\mathcal{W}},\bar{\bm{\mathcal W}}] = \frac{(-1)^s}{4} \int \rd^4x 
\rd^4 \q \, \bm{\mathcal{E}}	 \, \bm{\mathcal{W}}^{\a(2s+2)} \bm{\cW}_{\a(2s+2)}+ \text{c.c.}
\label{D.8}
\eea

\begin{footnotesize}

\end{footnotesize}


\end{document}